# A peridynamic theory for linear elastic shells

*Shubhankar Roy Chowdhury[1], Pranesh Roy[1], Debasish Roy[1,*] and J N Reddy[2]*

[1]Computational Mechanics Lab, Department of Civil Engineering, Indian Institute of Science, Bangalore 560012, India

[2]Advanced Computational Mechanics Lab, Department of Mechanical Engineering, Texas A&M University, College Station, Texas 77843-3123

(*Corresponding author; email: royd@civil.iisc.ernet.in)

**Abstract**

*A state-based peridynamic formulation for linear elastic shells is presented. The emphasis is on introducing, possibly for the first time, a general surface based peridynamic model to represent the deformation characteristics of structures that have one physical dimension much smaller than the other two. A new notion of curved bonds is exploited to cater for force transfer between the peridynamic particles describing the shell. Starting with the three dimensional force and deformation states, appropriate surface based force, moment and several deformation states are arrived at. Upon application on the curved bonds, such states beget the necessary force and deformation vectors governing the motion of the shell. Correctness of our proposal on the peridynamic shell theory is numerically assessed against static deformation of spherical and cylindrical shells and flat plates.*

**Keywords**: peridynamic shell; curved bond; surface based state; constitutive correspondence;

**1. Introduction**

Great leaps in computing power over the last few decades have enabled researchers to undertake numerical simulations of myriad complex, multi-scale and multi-physics problems of interest in continuum solid mechanics that were erstwhile unsolvable. Numerical techniques such as the finite element method (FEM) as well as mesh-free discretization schemes, e.g., smoothed particle hydrodynamics (SPH), reproducing kernel particle method (RKPM), moving least square Petrov-Galerkin method (MLPG) etc. have provided the necessary formalism in obtaining finite dimensional approximations to solutions of such initial boundary value problems (IBVP) that describe, often in terms of partial differential equations (PDEs), the laws of motion. Although PDE-centric formulations have been successful in modelling phenomena ranging from elastic response of solids to more complicated problems involving, for instance, propagating discontinuities in dynamic crack propagation, void growth and contact mechanics, its formal mathematical structure does not provide an ideal setup for applications to scenarios that must deal with evolution of discontinuities, e.g. crack nucleation and growth and other such problems in continuum damage mechanics. The partial derivatives in PDEs are not defined in the classical sense on the lines/surfaces of discontinuities and no valid diffeomorphism is at hand to relate the deformed and reference configurations. Computational methods for solving such problems using the PDE-based theory either require a redefinition of the object manifold so that discontinuities lie on the boundary or some special treatment to define spatial derivatives of field variables on a cracked surface (Bittencourt *et al*., 1996, Belytschko and Black, 1999, Areias and Belytschko, 2005).

Recently Silling (2000) introduced a reformulation of continuum theory, the peridynamics (PD), which can, by design, treat discontinuities. The primary feature that enables PD to deal with spontaneous emergence and propagation of discontinuity in solids, is the representation of equations of motion in integro-differential form, and not by PDEs. The integro-differential format of the governing equations relaxes, to a significant extent, the smoothness requirement of the deformation field, thereby allowing for discontinuities as long as the spatial integrals remain Riemann integrable. PD equations of motion are based on a model of internal forces that the material points exert on each other over finite distances. Such finite distance interactions lend a non-local character to the formulation and allow for length scale effects arising from the action at a distance. This inherent non-local feature in PD theory is useful in modelling a broad class of non-classical phenomena. An initial formulation, the bond-based PD, considered internal forces as a network of interacting pairs like springs, i.e. it described spring-like interactions via pair potentials. The interaction of a material particle with its surroundings is restricted to a finite neighborhood and is denoted as the horizon of the particle. Such pair-wise interaction however led to an over-simplification of the model and in particular resulted in an effective Poisson's ratio of 1/4 in case of linear isotropic elastic materials. This limitation has been overcome through a generalization of the PD model, the state-based PD (Silling *et al*., 2007). According to the state-based PD philosophy, the bond force between two interacting particles is no longer governed by a central potential independent of the behavior of other bonds; instead it is determined by the collective deformations of bonds within the horizon of a material particle. This version of the PD theory eliminates the restriction on Poisson's ratio and is applicable over the entire permissible range. Even though the PD has many attractive features, the scarcity of strictly PD-based material constitutive models tends to limit its applicability. A remedy to this is however proposed using a constitutive correspondence framework (Silling *et al*., 2007), enabling the use of classical material models in a PD formulation.

These features of the PD have attracted interest in solving solid mechanics problems especially those involving material damage. Most such attempts deal with the full-blown 3D model of continua, whilst a few rest consider the in-plane response within plane stress or plane strain type material modelling. Even though examples of structures resisting transverse deformation with one dimension (e.g. the thickness) significantly smaller than the other two are aplenty (e.g. aircraft fuselage, ship hull, pressure vessel, roofs of civil structures, turbine blades etc.), very few attempts in the PD literature are available that exploit the possibility of efficiently modelling such 3D bodies in terms of locally 2D equations of motion. To cite some instances of dimensionally reduced PD models, we refer to Silling *et al*. (2003) for 1D bar formulation, Chowdhury *et al*. (2015) and O'Grady and Foster, 2014 for 1D beams, Silling and Bobaru, (2005) for 2D membranes, Taylor and Steigmann (2013) and O'Grady and Foster (2014) for plates and flat shells. Such studies are useful to relieve the extensive computational overhead generally associated with the discretization of a 3D model. For instance, if the so-called thickness dimension is small, use of the 3D model would typically demand a rather fine discretization in the through-thickness direction en route to an accurate representation of the resistance to bending. This may need prohibitively expensive computational effort. The 2D formulations address this issue by an analytical accounting of the stress and deformation fields along the thickness direction and thus avoid through-thickness discretization. Studies by Taylor and Steigmann (2013) and O'Grady and Foster (2014) reflect a few such efforts, with the former reducing a 3D bond-based PD formulation to 2D in order to model bending characteristic of plates and the latter deriving a state-based PD model for plates and flat shells. While the bond based plate formulation of Taylor and Steigmann suffers from the usual limitation of Poisson's ratio being reduced to 1/4, O'Grady and Foster's state-based model is applicable for the entire permissible range of Poisson's ratio. Even though the two models noted above can analyze transverse deformation of flat structures, deformation of a 'thin' 3D body that may be described with reference to a curved base surface cannot be analyzed by them. Owing to the curvature of the surface, analysis of such structures is more complicated than that of flat structures, as transverse bending effects generally get coupled with stretching. Moreover a flat plate model may always be recovered as a limiting case of a curved shell model. To the best of the authors' knowledge, there exists no generic surface-based PD formulation to deal with such problems. In contrast, the literature on classical continuum mechanics is rich with an abundance of shell models. In classical shell modelling, kinematics and kinetics of the 3D body are described through tensorial quantities defined over a base surface, typically the mid surface. The governing equations of motion and constitutive relations for a shell, in terms of these surface based tensorial quantities, may be arrived at via through-thickness integration of the 3D equations of motion and constitutions respectively. Shell equations so obtained include only two independent surface coordinates vis-a-vis three independent ones in the 3D model. Strain measures referred to the base surface may be obtained from the power balance equation as quantities conjugate to shell stress measures. Assumed variation of the displacement field along the thickness direction is made use of in identifying the appropriate shell strains (Reissner, 1941, Naghdi, 1973). Variational asymptotic method provides an alternative route to deriving the classical shell equations from 3D (Berdichevskii, 1979).

In the present work, a surface based PD formulation, possibly the first of its kind and applicable for general curved shells, is set forth. A non-ordinary state-based approach is adopted. The 3D state-based PD equations are reduced to their surface representations by defining new force and moment state fields obtainable from 3D force states upon appropriate integration over the thickness direction. The proposed set of equations that describes the motion of shell is shown to satisfy the global requirements of linear and angular momenta balance for the 3D body. New deformation state fields referred to the base surface are identified from the energy balance equation. These states appear as conjugate quantities to the force and moment states in internal energy expression. A new notion of curved bonds replacing straight bonds in the standard PD theory is introduced. The curved bonds facilitate transfer of force and moment between PD particles. Also the 'size' of the horizon of a particle, which is of curved surface geometry, is decided by fixing a maximum length of curved bonds within it. In order to write down the material model, a constitutive correspondence route is followed. The necessary classical constitutive relations required for the correspondence are derived and given in Appendix I. Presently the scope of the modelling is restricted to small elastic deformation of thin shells. A few numerical tests concerning static deformation under transverse load are undertaken for purposes of validation. The numerical illustrations include problems on singly and doubly curved shells and as a special case a plate problem.

The rest of the paper is organized as follows. Section 2 briefly describes the state based PD theory and also gives a short account of the classical shell formulation with necessary details on the theory of surfaces. Section 3 reports on a systematic derivation of the PD shell formulation, which includes shell equations of motion, balance of energy and constitutive relations. This is followed by numerical illustrations and a few concluding remarks in Sections 4 and 5 respectively.

**2. State based PD and classical shell modelling**

For completeness, the state based PD theory along with the constitutive correspondence is briefly reviewed in this section. To aid in the derivation of PD shell equations, a summary of the classical shell equations and certain notions in differential geometry necessary for the development are also included here.

*2.1 State based PD*

Following the approach in (Silling *et al.*, 2007), a brief account of the state-based PD theory is presented below. PD, a non-local continuum theory allowing finite distance interaction between material particles, describes the dynamics of a body occupying a region $\mathcal{B}_o \subset \mathbb{R}^3$ in its reference configuration and $\mathcal{B}_t \subset \mathbb{R}^3$ in the current configuration. Figure 1 shows a schematic of the body. The vector $\xi = X'-X$ between a material point $X \in \mathcal{B}_o$ and its neighbour $X' \in \mathcal{B}_o$ is referred to as the *bond* between them. The bond vector gets deformed under the deformation map $\chi : \mathcal{B}_o \to \mathcal{B}_t$ and the deformed bond is given by the deformation vector state $\underline{Y}$ (see Silling *et al.* 2007, for a precise definition of states),

$$\underline{Y}[X]\langle\xi\rangle = y' - y = \chi(X') - \chi(X). \tag{1}$$

The finite distance over which the interaction of a point $X$ with its neighbor is to be considered is given by its *horizon* $\mathcal{H}$, defined using the family of bonds associated with the material point $X$ as $\mathcal{H}(X) = \{\xi \in \mathbb{R}^3 \mid (\xi + X) \in \mathcal{B}_0, |\xi| < \delta\}$, where $\delta > 0$ is the radius of the horizon.

The equations of motion in the state based PD are of the following integro-differential form.

$$\rho(X)\ddot{y}(X,t) = \int_{\mathcal{H}(X)} \{\underline{T}[X,t]\langle\xi\rangle - \underline{T}[X+\xi,t]\langle-\xi\rangle\} dV_{X'} + b(X,t), \tag{2}$$

where $\rho, \underline{\mathbf{T}}, \mathbf{b}$ are the mass density, long range internal force vector state and externally applied body force density respectively. Superimposed dots indicate material derivatives with respect to time. This equation has been shown in (Silling *et al.*, 2007) to satisfy the global balance of linear momentum, i.e. one which is valid for the entire body. Global conservation of angular momentum leads to the following restriction on the constitutive relation.

$$\int_{\mathcal{H}(X)} \underline{\mathbf{T}}[X,t]\langle\xi\rangle \times \underline{\mathbf{Y}}[X,t]\langle\xi\rangle dV_{X'} = 0, \quad \forall X \in \mathcal{B}_0 \tag{3}$$

In order to incorporate within the PD setup the rich repertoire of material models in classical solid mechanics, which the PD lacks, Silling *et al.* (2007) have proposed a way of constitutive correspondence. This correspondence is obtained by a non-local definition of the deformation gradient in terms of the deformation state and then equating the PD-based internal energy to the classical one, written in terms of the non-local deformation gradient, for the same deformation. The correspondence relations are as shown below.

$$\overline{F}(\underline{\mathbf{Y}}) = \left[\int_{\mathcal{H}} \omega(|\xi|)(\underline{\mathbf{Y}}\langle\xi\rangle \otimes \xi) dV_{X'}\right] \overline{K}^{-1} \tag{4}$$

$$\overline{K} = \int_{\mathcal{H}} \omega(|\xi|)(\xi \otimes \xi) dV_{X'} \tag{5}$$

$$\underline{\mathbf{T}}\langle\xi\rangle = \omega(|\xi|)\overline{P}\overline{K}^{-1}\xi \tag{6}$$

Here $\overline{F} \in \mathbb{R}^3 \times \mathbb{R}^3$ is the non-local deformation gradient tensor, $\overline{K} \in \mathbb{R}^3 \times \mathbb{R}^3$ a non-local shape tensor, $\omega$ a non-negative scalar influence function ($\omega(|\xi|) > 0, \forall \xi \in \mathcal{H}$) and $\overline{P} = \tilde{P}(\overline{F})$ the first Piola-Kirchhoff stress obtained from the classical constitutive relation for $\tilde{P}$ written in terms of the non-local deformation gradient $\overline{F}$. It has also been shown (Silling *et al.*, 2007) that such correspondence satisfies the restriction given in Eq. (3) and hence the conservation of angular momentum for the body is ensured.

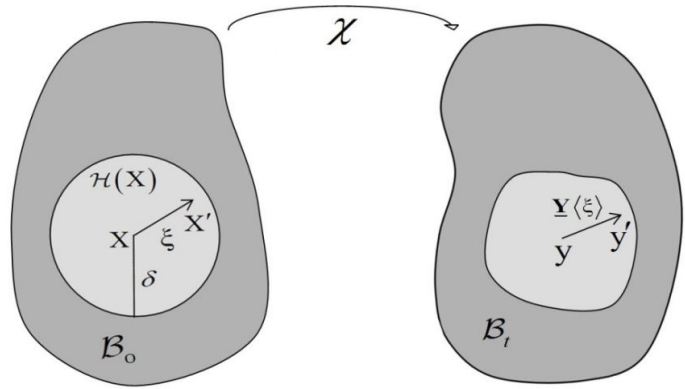

Figure 1. Schematic PD body in the reference and current configurations

*2.1 Classical shell modelling*

An ideal interpretation of a shell would be as a 3D body bounded by two curved surfaces, the distance between which - namely the thickness of the shell - is small in comparison with the other dimensions of the body. In other words, a shell resembles a curved surface having a thickness. The mathematical theory of shells seeks an approximation to describe the response of the 3D body using a local 2D model given on the surface and the formulation takes the form of a set of PDEs representing an IBVP with the surface as its domain. Typically it is the middle surface that is chosen as the base surface to which the 2D model is referred. An essential step in any shell theory is to eliminate the third spatial coordinate, namely the one along the thickness direction, to obtain the surface based 2D equations. There are a few approaches that may be followed to reach this goal. One of them assumes a priori the form of dependence of the shell displacement and stress fields on the thickness coordinate aiding in the integration of the 3D equations along the thickness. This type of schemes may be referred to as hypothesis method. The Kirchhoff-Love theory of shell is an example of such method. Another formulation involves asymptotic expansion of the 3D solution with respect to a small parameter that subsequently derives a 2D model. It may be noted that there are other ways of deriving 2D shell equations without recourse to 3D elasticity, wherein equations of motions for the surface are obtained by directly making use of the fundamental laws of mechanics on the surface.

For purposes of the present article, we will consider the hypothesis method in our derivation of the shell equations. In view of the tensorial quantities being functions of curvilinear surface coordinates in a shell model, the theory of surfaces from differential geometry is an integral aspect of it. In this section we outline a model for thin linear elastic shells, assuming small deformation. Details of derivation of this classical shell equation along with a brief description of surface geometry are given in Appendix I.

The surface based 2D equations of motion for the shell are given by

$$\tilde{\nabla} \cdot \mathbf{S} + \mathbf{q} = \rho h \ddot{\mathbf{u}}_S, \qquad \tilde{\nabla} \cdot \mathbf{M} + \mathbf{S}_\times + \mathbf{m} = \frac{\rho h^3}{12} \ddot{\boldsymbol{\theta}} \qquad (7)$$

where $\mathbf{S}$ and $\mathbf{M}$ are the stress and moment tensors defined on the base surface of shell and $\mathbf{S}_\times$ is the vectorial invariant of $\mathbf{S}$. $\mathbf{q}$ and $\mathbf{m}$ are distributed force and moment type quantities expressed as force or moment per unit area. $\mathbf{u}_S = \tilde{\mathbf{u}} + w\mathbf{n}$ represents the displacement field and $\tilde{\boldsymbol{\theta}}$ the rotation field of the base surface of the shell. $\tilde{\nabla}$ is the gradient operator defined for the surface. The middot · denotes the scalar product and superimposed dots indicate material derivatives with respect to time. For a more detailed exposition and definitions of some of these terms and symbols, see Appendix I.

The shell kinematic relations are given though the surface strain and wryness tensor,

$$\boldsymbol{\gamma} = \left(\tilde{\nabla}(\tilde{\mathbf{u}} + w\mathbf{n})\right)^T - \mathbf{n} \otimes \mathbf{n} \times \tilde{\boldsymbol{\theta}}, \qquad \boldsymbol{\kappa} = \left(\tilde{\nabla}\tilde{\boldsymbol{\theta}}\right)^T \tag{8}$$

where $\mathbf{n}$ denotes the unit normal to the base surface, $\otimes$ is the symbol for tensor product and $\times$ denotes cross product. For a shell made of linear elastic isotropic material and satisfying $\|h\mathbf{B}\| \ll 1$, the reduced constitutive relations in terms of the surface stress and strain tensors may be written as follows.

$$\mathbf{S} = \lambda h \operatorname{tr}\left(\mathbf{A} \cdot \frac{\boldsymbol{\gamma} + \boldsymbol{\gamma}^T}{2}\right)\mathbf{A} + 2\mu h \left(\mathbf{A} \cdot \frac{\boldsymbol{\gamma} + \boldsymbol{\gamma}^T}{2}\right) \tag{9}$$

$$\mathbf{M} = \frac{\lambda h^3}{12}\left((\mathbf{A}.\boldsymbol{\kappa})^T - \mathbf{A}.\boldsymbol{\kappa}\right) + \frac{\lambda h^3}{6}\left((\mathbf{A}.\boldsymbol{\kappa})^T - \frac{1}{2}\operatorname{tr}\left((\mathbf{A}.\boldsymbol{\kappa})^T\right)\mathbf{A}\right) \tag{10}$$

Here $\lambda$ and $\mu$ are the Lamé parameters, $h$ is the thickness of shell, $\mathbf{A}$ is the metric tensor and $\mathbf{B}$ the curvature tensor defined for the base surface. See Appendix I for definitions of metric tensor $\mathbf{A}$ and curvature tensor $\mathbf{B}$.

## 3. PD Shell formulation

In order to predict the deformation characteristics of shell, one may attempt to solve the 3D equations of motions given in Eqn. (2) with a very fine discretization along the thickness direction. However the prohibitive computational demand and possible numerical ill conditioning that it associates, suggest a more expedient route through a 2D surface based PD model capable of fairly representing the shell dynamics. In this section we obtain the state-based 2D PD equations of motion. The aim here is to represent the 3D body of a shell via a single surface consisting of PD particles interacting with each other. As a first step, we alter the notion of PD bonds originally conceived as straight lines joining two PD particles. Instead of straight lines, we identify specific curved lines to be the bonds between PD particles on the base surface. A modified definition of the horizon of a point, which lies on the base surface, also follows from these curved bonds. A curved bond $\tilde{\boldsymbol{\xi}}$ from a point of position $\boldsymbol{\rho}$ to a neighbour (say $\boldsymbol{\rho}'$ is the position vector of the neighbour) is defined via the geodesic curve between them. The modified horizon of $\boldsymbol{\rho}$ denoted by $\tilde{\mathcal{H}}(\boldsymbol{\rho})$ may be defined as the family of the curved bonds $\{\tilde{\boldsymbol{\xi}}\}$ from $\boldsymbol{\rho}$ such that $len(\tilde{\boldsymbol{\xi}}) \leq \delta$ where $len(\cdot)$ is the length operator which returns the geodesic distance between the end points of the curved bond (see Figure 2(b)). We have used the position vector itself to refer to the associated points and also in the following sections, wherever convenient, we would continue to use a similar convention.

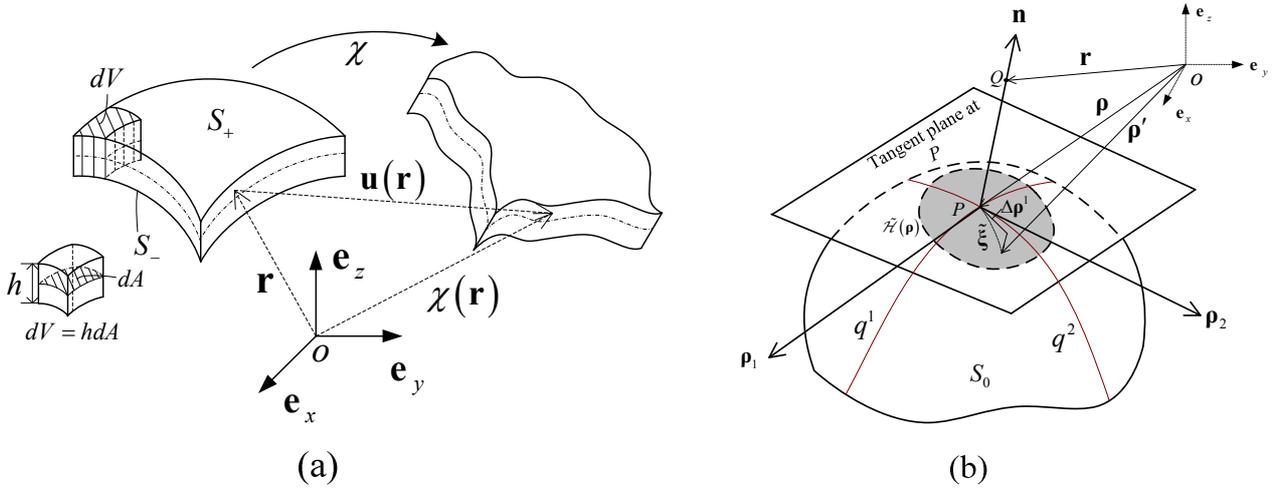

Figure 2. (a) Schematic of shell in the reference and current configurations; (b) Base surface, curved bond and horizon

*3.1. Kinematics*

Let a point $Q$ in the 3D body of the shell, given by its position vector $\mathbf{r}$, gets deformed under the deformation mapping $\chi$ to $\chi(\mathbf{r})$. The corresponding displacement vector $\mathbf{u}(\mathbf{r})$ is hypothesized to have the following form.

$$\mathbf{u}(\mathbf{r}) = \mathbf{u}_S(\boldsymbol{\rho}) + \tilde{\boldsymbol{\theta}}(\boldsymbol{\rho}) \times z\mathbf{n} \qquad (11)$$

Here $\mathbf{u}_S(\boldsymbol{\rho})$ and $\tilde{\boldsymbol{\theta}}(\boldsymbol{\rho})$ are respectively the deformation vector and the rotation pseudo-vector of the base surface of the shell at $P$, which is the base point of the normal $\mathbf{n}$ that passes through $Q$ (see Figure 2(b)). $z\mathbf{n}$ represents the vector $\overrightarrow{PQ}$. Note that $\mathbf{r} = \boldsymbol{\rho} + z\mathbf{n}$.

For a kinematical description of the shell, we need to define a few vector states defined over the base surface as listed below.

Relative surface displacement state, $\underline{\mathbf{U}}[\boldsymbol{\rho}]\langle\tilde{\boldsymbol{\xi}}\rangle = \mathbf{u}'_S - \mathbf{u}_S = (\chi(\boldsymbol{\rho}') - \boldsymbol{\rho}') - (\chi(\boldsymbol{\rho}) - \boldsymbol{\rho})$ (12)

Relative surface rotation state, $\underline{\boldsymbol{\Theta}}[\boldsymbol{\rho}]\langle\tilde{\boldsymbol{\xi}}\rangle = \tilde{\boldsymbol{\theta}}' - \tilde{\boldsymbol{\theta}} = \tilde{\boldsymbol{\theta}}(\boldsymbol{\rho}') - \tilde{\boldsymbol{\theta}}(\boldsymbol{\rho})$ (13)

Averaged surface rotation state, $\underline{\hat{\boldsymbol{\Theta}}}[\boldsymbol{\rho}]\langle\tilde{\boldsymbol{\xi}}\rangle = \frac{1}{2}(\tilde{\boldsymbol{\theta}}' + \tilde{\boldsymbol{\theta}})$ (14)

The approximations involved in the vector operations above on rotation pseudo-vectors at different base points must be qualified against the 'smallness' of either the horizon or possibly the shell curvature.

*3.2 Equations of motion*

We postulate the following two sets of equations to describe the motion of PD shell.

$$\rho h \ddot{\mathbf{u}}_S = \int_{\tilde{\mathcal{H}}(\boldsymbol{\rho})} \{\underline{\mathbf{S}}[\boldsymbol{\rho}]\langle\tilde{\boldsymbol{\xi}}\rangle - \underline{\mathbf{S}}[\boldsymbol{\rho}']\langle-\tilde{\boldsymbol{\xi}}\rangle\} \, dA_{\boldsymbol{\rho}'} + \mathbf{b}_S \quad (15)$$

$$\frac{\rho h^3}{12} \ddot{\tilde{\boldsymbol{\theta}}} = \int_{\tilde{\mathcal{H}}(\boldsymbol{\rho})} \{\underline{\boldsymbol{\mu}}[\boldsymbol{\rho}]\langle\tilde{\boldsymbol{\xi}}\rangle - \underline{\boldsymbol{\mu}}[\boldsymbol{\rho}']\langle-\tilde{\boldsymbol{\xi}}\rangle\} \, dA_{\boldsymbol{\rho}'} + \frac{1}{2}\int_{\tilde{\mathcal{H}}(\boldsymbol{\rho})} (\boldsymbol{\rho}' - \boldsymbol{\rho}) \times \{\underline{\mathbf{S}}[\boldsymbol{\rho}]\langle\tilde{\boldsymbol{\xi}}\rangle - \underline{\mathbf{S}}[\boldsymbol{\rho}']\langle-\tilde{\boldsymbol{\xi}}\rangle\} \, dA_{\boldsymbol{\rho}'} + \mathbf{l}_S \quad (16)$$

In the above equations $\underline{\mathbf{S}}$ and $\underline{\boldsymbol{\mu}}$ are respectively the PD force and moment state fields defined over the base surface. $dA_{\boldsymbol{\rho}}$ is the area element on the base surface at $\boldsymbol{\rho}$. In terms of the local reciprocal basis vectors, the 3D force state $\underline{\mathbf{T}}$ at $\mathbf{r}$ takes the form $\underline{\mathbf{T}}[\mathbf{r}]\langle\mathbf{r}' - \mathbf{r}\rangle = T_1\mathbf{r}^1 + T_2\mathbf{r}^2 + T_3\mathbf{n}$ where $\{T_i\}$ are functions of both $\mathbf{r}$ and $\mathbf{r}'$, the latter denoting the position vector of a neighbouring material point. Similarly $\underline{\mathbf{S}}[\boldsymbol{\rho}]\langle\tilde{\boldsymbol{\xi}}\rangle = S_1\boldsymbol{\rho}^1 + S_2\boldsymbol{\rho}^2 + S_3\mathbf{n}$ where $\{S_i\}$ depend on $\boldsymbol{\rho}$ and $\boldsymbol{\rho}'$. Note that $\boldsymbol{\rho}'$ corresponds to $\mathbf{r}'$ in the same way as $\boldsymbol{\rho}$ to $\mathbf{r}$. For details about the bases, see section A1 in Appendix I. The covariant components $S_i$ are defined in terms of $T_i$ as

$$S_\gamma = \iint_{z\,z'} (\mathbf{A} - z\mathbf{B})^{-1\alpha}_{\gamma} T_\alpha G' dz' G dz, \qquad S_3 = \iint_{z\,z'} T_3 G' dz' G dz \quad (17)$$

In this expression and what follows, summation convention over repeated indices (Greek alphabets; e.g., $\alpha, \gamma$) is adopted and they can take values 1 and 2. In Eqn. (17) $G = \det(\mathbf{A} - z\mathbf{B})$. In direct (coordinate-free) notation, Eqn. (17) is

$$\underline{\mathbf{S}}[\boldsymbol{\rho}]\langle\tilde{\boldsymbol{\xi}}\rangle = \int_z \int_{z'} \underline{\mathbf{T}}[\mathbf{r}]\langle\mathbf{r}'-\mathbf{r}\rangle G'G dz'dz \qquad (18)$$

In terms of $\underline{\mathbf{T}}$, the moment state $\underline{\boldsymbol{\mu}}$ is written as

$$\underline{\boldsymbol{\mu}}[\boldsymbol{\rho}]\langle\tilde{\boldsymbol{\xi}}\rangle = \frac{1}{2}\int_z \int_{z'} (z'\mathbf{n}' + z\mathbf{n}) \times \underline{\mathbf{T}}[\mathbf{r}]\langle\mathbf{r}'-\mathbf{r}\rangle G'G dz'dz \qquad (19)$$

The distributed force $\mathbf{b}_S$ and moment $\mathbf{l}_S$ on the surface are given in terms of the body force $\mathbf{b}$ as

$$\mathbf{b}_S = \int_{-h/2}^{h/2} \mathbf{b} G dz \quad \text{and} \quad \mathbf{l}_S = \int_{-h/2}^{h/2} z\mathbf{n} \times \mathbf{b} G dz. \qquad (20)$$

For this formulation the mass density $\rho$ is considered not to vary along the thickness; hence it is independent of $z$.

In order to meet the physical principles, it is necessary that the states $\underline{\mathbf{S}}$ and $\underline{\boldsymbol{\mu}}$ satisfy the balance of linear and angular momenta for the 3D body $\mathcal{B}$ of the shell. Mathematically, this requires the satisfaction of the following two conditions.

$$\int_{\mathcal{B}} \rho(\mathbf{r})\ddot{\mathbf{u}}(\mathbf{r},t) dV_{\mathbf{r}} = \int_{\mathcal{B}} \mathbf{b}(\mathbf{r},t) dV_{\mathbf{r}} \qquad (21)$$

$$\int_{\mathcal{B}} \chi(\mathbf{r}) \times \rho(\mathbf{r})\ddot{\mathbf{u}}(\mathbf{r},t) dV_{\mathbf{r}} = \int_{\mathcal{B}} \chi(\mathbf{r}) \times \mathbf{b}(\mathbf{r},t) dV_{\mathbf{r}} \qquad (22)$$

For small deformation, Eqn. (22) may be approximated as

$$\int_{\mathcal{B}} \mathbf{r} \times \rho(\mathbf{r})\ddot{\mathbf{u}}(\mathbf{r},t) dV_{\mathbf{r}} = \int_{\mathcal{B}} \mathbf{r} \times \mathbf{b}(\mathbf{r},t) dV_{\mathbf{r}} \qquad (23)$$

**Proposition 3.1** *Let the base surface $S_0$ of the shell $\mathcal{B}$ be subjected to distributed force and moment density (per unit area) $\mathbf{b}_S$ and $\mathbf{l}_S$ as defined in Eqn. (20). With $\underline{\mathbf{S}}$ and $\underline{\boldsymbol{\mu}}$ respectively denoting the internal force and moment vector states on the base surface, if Eqn. (15) holds in $S_0$ then balance of linear momentum for $\mathcal{B}$ is satisfied, i.e., Eqn. (21) holds.*

*Proof:* Integrate Eqn. (15) over the base surface area $\mathcal{A}$. Since no interaction exists beyond the horizon, extend the inner integral (the one appearing on the right hand side of Eqn. (15)) to $\mathcal{A}$ leading to $\int_{\mathcal{A}} \rho h \ddot{\mathbf{u}}_S dA_{\boldsymbol{\rho}} = \int_{\mathcal{A}} \int_{\mathcal{A}} \{\underline{\mathbf{S}}[\boldsymbol{\rho}]\langle\tilde{\boldsymbol{\xi}}\rangle - \underline{\mathbf{S}}[\boldsymbol{\rho}']\langle-\tilde{\boldsymbol{\xi}}\rangle\} dA_{\boldsymbol{\rho}'}dA_{\boldsymbol{\rho}} + \int_{\mathcal{A}} \mathbf{b}_S dA_{\boldsymbol{\rho}}$.

Substitute for $\underline{\mathbf{S}}$ from Eqn. (18) and for $\mathbf{b}_S$ from Eqn. (20) to recast the above equation as

$$\int_{\mathcal{A}} \int_{-h/2}^{h/2} \rho\left(\ddot{\mathbf{u}}_S + \ddot{\boldsymbol{\theta}} \times z\mathbf{n}\right) G dz dA_\rho$$
$$= \int_{\mathcal{A}} \int_{\mathcal{A}} \int_{-h/2}^{h/2} \int_{-h/2}^{h/2} \left\{\underline{\mathbf{T}}[\mathbf{r}]\langle\mathbf{r}'-\mathbf{r}\rangle - \underline{\mathbf{T}}[\mathbf{r}']\langle\mathbf{r}-\mathbf{r}'\rangle\right\} G' G dz' dz dA_{\rho'} dA_\rho + \int_{\mathcal{A}} \int_{-h/2}^{h/2} \mathbf{b} G dz dA_\rho \quad (24)$$

The left hand side follows upon replacing $\rho h \ddot{\mathbf{u}}_S$ by $\int_{-h/2}^{h/2} \rho\left(\ddot{\mathbf{u}}_S + \ddot{\boldsymbol{\theta}} \times z\mathbf{n}\right) dz$. Identifying $G dz dA_\rho$ as the volume element $dV_\rho$, Eqn. (24) may be written as

$$\int_{\mathcal{B}} \rho \ddot{\mathbf{u}} dV_\rho = \int_{\mathcal{B}} \int_{\mathcal{B}} \left\{\underline{\mathbf{T}}[\mathbf{r}]\langle\mathbf{r}'-\mathbf{r}\rangle - \underline{\mathbf{T}}[\mathbf{r}']\langle\mathbf{r}-\mathbf{r}'\rangle\right\} dV_{\rho'} dV_\rho + \int_{\mathcal{B}} \mathbf{b} dV_\rho$$

First term on the right hand side may be shown to vanish using a change of variables $\mathbf{r} \leftrightarrow \mathbf{r}'$ and the subsequent application of Fubini's theorem. Hence the proof follows. □

**Proposition 3.2** *With the same conditions as in Proposition 3.1, if Eqn. (16) along with Eqn. (15) hold in $S_0$, balance of angular momentum for $\mathcal{B}$ is satisfied, i.e., Eqn. (23) holds.*

*Proof:* On the lines of the previous proof, integrating Eqn. (16) over $\mathcal{A}$ and extending the inner integrals to $\mathcal{A}$, we get

$$\int_{\mathcal{B}} z\mathbf{n} \times \rho\ddot{\mathbf{u}} dV_\rho = \frac{1}{2} \int_{\mathcal{B}} \int_{\mathcal{B}} (z'\mathbf{n}' + z\mathbf{n}) \times \left\{\underline{\mathbf{T}}[\mathbf{r}]\langle\mathbf{r}'-\mathbf{r}\rangle - \underline{\mathbf{T}}[\mathbf{r}']\langle\mathbf{r}-\mathbf{r}'\rangle\right\} dV_{\rho'} dV_\rho$$
$$+ \frac{1}{2}\int_{B}\int_{\mathcal{B}}(\boldsymbol{\rho}'-\boldsymbol{\rho})\times\left\{\underline{\mathbf{T}}[\mathbf{r}]\langle\mathbf{r}'-\mathbf{r}\rangle - \underline{\mathbf{T}}[\mathbf{r}']\langle\mathbf{r}-\mathbf{r}'\rangle\right\} dV_{\rho'} dV_\rho + \int_{\mathcal{B}} z\mathbf{n} \times \mathbf{b} dV_\rho \quad (25)$$

Here $\int_{-h/2}^{h/2} \rho z\mathbf{n} \times \ddot{\mathbf{u}} G dz$ is substituted for $\frac{\rho h^3}{12}\ddot{\boldsymbol{\theta}}$ and other substitutions follows form Eqn. (19) and (20).

Rearranging Eqn. (25), and using $\mathbf{r} = \boldsymbol{\rho} + z\mathbf{n}$, we have

$$\int_{\mathcal{B}} \mathbf{r} \times (\rho\ddot{\mathbf{u}} - \mathbf{b}) dV_\rho = \frac{1}{2}\int_{\mathcal{B}}\int_{\mathcal{B}}(\mathbf{r}'+\mathbf{r})\times\left\{\underline{\mathbf{T}}[\mathbf{r}]\langle\mathbf{r}'-\mathbf{r}\rangle - \underline{\mathbf{T}}[\mathbf{r}']\langle\mathbf{r}-\mathbf{r}'\rangle\right\} dV_{\rho'} dV_\rho$$
$$+ \int_{B}\boldsymbol{\rho}\times\left((\rho\ddot{\mathbf{u}}-\mathbf{b}) - \int_{\mathcal{B}}\left\{\underline{\mathbf{T}}[\mathbf{r}]\langle\mathbf{r}'-\mathbf{r}\rangle - \underline{\mathbf{T}}[\mathbf{r}']\langle\mathbf{r}-\mathbf{r}'\rangle\right\} dV_{\rho'}\right) dV_\rho$$

Both terms on the RHS are identically zero, first of which again follows from a change of variables $\mathbf{r} \leftrightarrow \mathbf{r}'$ followed by application of Fubini's theorem and the second holds because of Eqn. (15). This may be seen by writing this term in the following alternative form.

$$\int_{\mathcal{A}} \boldsymbol{\rho} \times \left( \int_{-h/2}^{h/2} (\rho \ddot{\mathbf{u}} - \mathbf{b}) G dz - \int_{\mathcal{A}} \int_{-h/2}^{h/2} \int_{-h/2}^{h/2} \{\underline{\mathbf{T}}[\mathbf{r}]\langle \mathbf{r}' - \mathbf{r}\rangle - \underline{\mathbf{T}}[\mathbf{r}']\langle \mathbf{r} - \mathbf{r}'\rangle\} G' dz' G dz dA_{\rho'} \right) dA_{\rho}$$

$$= \int_{\mathcal{A}} \boldsymbol{\rho} \times \left( (\rho \ddot{\mathbf{u}}_S - \mathbf{b}_S) - \int_{\mathcal{A}} \{\underline{\mathbf{S}}[\boldsymbol{\rho}]\langle \boldsymbol{\xi}\rangle - \underline{\mathbf{S}}[\boldsymbol{\rho}']\langle \boldsymbol{\xi}'\rangle\} dA_{\rho'} \right) dA_{\rho} = \mathbf{0}$$

Hence the claim follows. □

*3.3 Energy balance and constitutive relations*

The present focus will only be on the mechanical version of energy balance without considerations of any heat source or flux. Scalar products of both sides of Eqn. (15) with surface velocity $\dot{\mathbf{u}}_S$ and Eqn. (16) with $\dot{\tilde{\boldsymbol{\theta}}}$, their subsequent addition and integration over a finite sub-region $\mathcal{P} \subset \mathcal{A}$ result in the following.

$$\frac{d}{dt}\int_{\mathcal{P}} \frac{\rho h \dot{\mathbf{u}}_S \cdot \dot{\mathbf{u}}_S}{2} dA_{\rho} + \frac{d}{dt}\int_{\mathcal{P}} \frac{\rho h^3 \dot{\tilde{\boldsymbol{\theta}}} \cdot \dot{\tilde{\boldsymbol{\theta}}}}{24} dA_{\rho} = \int_{\mathcal{P}} \int_{\mathcal{A}} \{\underline{\mathbf{S}} - \underline{\mathbf{S}}'\} \cdot \dot{\mathbf{u}}_S \, dA_{\rho'} dA_{\rho} + \int_{\mathcal{P}} \int_{\mathcal{A}} \{\underline{\boldsymbol{\mu}} - \underline{\boldsymbol{\mu}}'\} \cdot \dot{\tilde{\boldsymbol{\theta}}} \, dA_{\rho'} dA_{\rho}$$
$$+ \frac{1}{2}\int_{\mathcal{P}} \int_{\mathcal{A}} (\boldsymbol{\rho}' - \boldsymbol{\rho}) \times \{\underline{\mathbf{S}} - \underline{\mathbf{S}}'\} \cdot \dot{\tilde{\boldsymbol{\theta}}} \, dA_{\rho'} dA_{\rho} + \int_{\mathcal{P}} \mathbf{b}_S \cdot \dot{\mathbf{u}}_S \, dA_{\rho} + \int_{\mathcal{P}} \mathbf{l}_S \cdot \dot{\tilde{\boldsymbol{\theta}}} \, dA_{\rho} \quad (26)$$

Note that all the inner integrals in the expression above have been extended to the whole base surface with the same argument of no interaction beyond the horizon as used earlier. For conciseness, Eqn. (26) is written with some notational abuse. By $\underline{\mathbf{S}}$ and $\underline{\mathbf{S}}'$ we mean $\underline{\mathbf{S}}[\boldsymbol{\rho}]\langle\tilde{\boldsymbol{\xi}}\rangle$ and $\underline{\mathbf{S}}[\boldsymbol{\rho}']\langle-\tilde{\boldsymbol{\xi}}\rangle$ respectively; similar statements hold for $\underline{\boldsymbol{\mu}}$ and $\underline{\boldsymbol{\mu}}'$. Eqn. (26) basically represents the power balance for a portion of the 3D shell (that encapsulates the region $\mathcal{P}$ of the base surface) written in terms of the states defined over the surface and consists of kinetic energy rate, absorbed and supplied power as may be seen in what follows.

A few identities needed for further simplifications are listed below.

$$\{\underline{\mathbf{S}}\langle\tilde{\boldsymbol{\xi}}\rangle - \underline{\mathbf{S}}'\langle-\tilde{\boldsymbol{\xi}}\rangle\} \cdot \dot{\mathbf{u}}_S = \left(\underline{\mathbf{S}}\langle\tilde{\boldsymbol{\xi}}\rangle \cdot \dot{\mathbf{u}}'_S - \underline{\mathbf{S}}'\langle-\tilde{\boldsymbol{\xi}}\rangle \cdot \dot{\mathbf{u}}_S\right) - \underline{\mathbf{S}}\langle\tilde{\boldsymbol{\xi}}\rangle \cdot (\dot{\mathbf{u}}'_S - \dot{\mathbf{u}}_S) \quad (27)$$

$$\frac{1}{2}(\boldsymbol{\rho}' - \boldsymbol{\rho}) \times \{\underline{\mathbf{S}}\langle\tilde{\boldsymbol{\xi}}\rangle - \underline{\mathbf{S}}'\langle-\tilde{\boldsymbol{\xi}}\rangle\} \cdot \dot{\tilde{\boldsymbol{\theta}}} = \left(\underline{\mathbf{S}}\langle\tilde{\boldsymbol{\xi}}\rangle \cdot \frac{\dot{\tilde{\boldsymbol{\theta}}}'}{2} \times (\boldsymbol{\rho} - \boldsymbol{\rho}') - \underline{\mathbf{S}}'\langle-\tilde{\boldsymbol{\xi}}\rangle \cdot \frac{\dot{\tilde{\boldsymbol{\theta}}}}{2} \times (\boldsymbol{\rho}' - \boldsymbol{\rho})\right) + \underline{\mathbf{S}}\langle\tilde{\boldsymbol{\xi}}\rangle \cdot \frac{\dot{\tilde{\boldsymbol{\theta}}} + \dot{\tilde{\boldsymbol{\theta}}}'}{2} \times (\boldsymbol{\rho}' - \boldsymbol{\rho}) \quad (28)$$

$$\{\underline{\boldsymbol{\mu}}\langle\tilde{\boldsymbol{\xi}}\rangle - \underline{\boldsymbol{\mu}}'\langle-\tilde{\boldsymbol{\xi}}\rangle\} \cdot \dot{\tilde{\boldsymbol{\theta}}} = \left(\underline{\boldsymbol{\mu}}\langle\tilde{\boldsymbol{\xi}}\rangle \cdot \dot{\tilde{\boldsymbol{\theta}}}' - \underline{\boldsymbol{\mu}}'\langle-\tilde{\boldsymbol{\xi}}\rangle \cdot \dot{\tilde{\boldsymbol{\theta}}}\right) - \underline{\boldsymbol{\mu}}\langle\tilde{\boldsymbol{\xi}}\rangle \cdot \left(\dot{\tilde{\boldsymbol{\theta}}}' - \dot{\tilde{\boldsymbol{\theta}}}\right) \quad (29)$$

Using these identities, each of the first three terms on the right hand side of Eqn. (26) could be split into two integrals as illustrated below.

$$\int_{\mathcal{P}}\int_{\mathcal{A}}\{\underline{\mathbf{S}}-\underline{\mathbf{S}}'\}\cdot\dot{\mathbf{u}}_S\,dA_{\rho'}dA_{\rho} = \int_{\mathcal{P}}\int_{\mathcal{A}}(\underline{\mathbf{S}}\cdot\dot{\mathbf{u}}'_S - \underline{\mathbf{S}}'\cdot\dot{\mathbf{u}}_S)dA_{\rho'}dA_{\rho} - \int_{\mathcal{P}}\int_{\mathcal{A}}\underline{\mathbf{S}}\cdot(\dot{\mathbf{u}}'_S - \dot{\mathbf{u}}_S)dA_{\rho'}dA_{\rho}$$

$$= \int_{\mathcal{P}}\int_{\mathcal{A}\backslash\mathcal{P}}(\underline{\mathbf{S}}\cdot\dot{\mathbf{u}}'_S - \underline{\mathbf{S}}'\cdot\dot{\mathbf{u}}_S)dA_{\rho'}dA_{\rho} - \int_{\mathcal{P}}\int_{\mathcal{A}}\underline{\mathbf{S}}\cdot(\dot{\mathbf{u}}'_S - \dot{\mathbf{u}}_S)dA_{\rho'}dA_{\rho}, \tag{30}$$

where the antisymmetry of the integrand in the first integral (on the right hand side) is used to obtain the last step. Similarly,

$$\frac{1}{2}\int_{\mathcal{P}}\int_{\mathcal{A}}(\boldsymbol{\rho}'-\boldsymbol{\rho})\times\{\underline{\mathbf{S}}-\underline{\mathbf{S}}'\}\cdot\dot{\tilde{\boldsymbol{\theta}}}\,dA_{\rho'}dA_{\rho}$$

$$= \int_{\mathcal{P}}\int_{\mathcal{A}\backslash\mathcal{P}}\left(\underline{\mathbf{S}}\cdot\frac{\dot{\tilde{\boldsymbol{\theta}}}'}{2}\times(\boldsymbol{\rho}-\boldsymbol{\rho}') - \underline{\mathbf{S}}'\cdot\frac{\dot{\tilde{\boldsymbol{\theta}}}}{2}\times(\boldsymbol{\rho}'-\boldsymbol{\rho})\right)dA_{\rho'}dA_{\rho} + \int_{\mathcal{P}}\int_{\mathcal{A}}\underline{\mathbf{S}}\cdot\frac{\dot{\tilde{\boldsymbol{\theta}}}+\dot{\tilde{\boldsymbol{\theta}}}'}{2}\times(\boldsymbol{\rho}'-\boldsymbol{\rho})dA_{\rho'}dA_{\rho} \tag{31}$$

and

$$\int_{\mathcal{P}}\int_{\mathcal{A}}\{\underline{\boldsymbol{\mu}}-\underline{\boldsymbol{\mu}}'\}\cdot\dot{\tilde{\boldsymbol{\theta}}}\,dA_{\rho'}dA_{\rho} = \int_{\mathcal{P}}\int_{\mathcal{A}\backslash\mathcal{P}}\left(\underline{\boldsymbol{\mu}}\cdot\dot{\tilde{\boldsymbol{\theta}}}' - \underline{\boldsymbol{\mu}}'\cdot\dot{\tilde{\boldsymbol{\theta}}}\right)dA_{\rho'}dA_{\rho} - \int_{\mathcal{P}}\int_{\mathcal{A}}\underline{\boldsymbol{\mu}}\cdot\left(\dot{\tilde{\boldsymbol{\theta}}}'-\dot{\tilde{\boldsymbol{\theta}}}\right)dA_{\rho'}dA_{\rho}. \tag{32}$$

Using Eqn. (30)-(32), Eqn. (26) could be written in the following form of power balance.

$$\dot{\mathcal{K}}(\mathcal{P}) + \mathcal{W}_{abs}(\mathcal{P}) = \mathcal{W}_{sup}(\mathcal{P}), \tag{33}$$

where $\mathcal{K}(\mathcal{P}) = \int_{\mathcal{P}}\frac{\rho h\dot{\mathbf{u}}_S\cdot\dot{\mathbf{u}}_S}{2}dA_{\rho} + \int_{\mathcal{P}}\frac{\rho h^3\dot{\tilde{\boldsymbol{\theta}}}\cdot\dot{\tilde{\boldsymbol{\theta}}}}{24}dA_{\rho}$ is the kinetic energy in $\mathcal{P}$. The absorbed power by $\mathcal{P}$ and supplied power to $\mathcal{P}$ are respectively given by

$$\mathcal{W}_{abs}(\mathcal{P}) = \int_{\mathcal{P}}\int_{\mathcal{A}}\underline{\mathbf{S}}\cdot\left\{(\dot{\mathbf{u}}'_S - \dot{\mathbf{u}}_S) - \frac{\dot{\tilde{\boldsymbol{\theta}}}+\dot{\tilde{\boldsymbol{\theta}}}'}{2}\times(\boldsymbol{\rho}'-\boldsymbol{\rho})\right\}dA_{\rho'}dA_{\rho} + \int_{\mathcal{P}}\int_{\mathcal{A}}\underline{\boldsymbol{\mu}}\cdot\left(\dot{\tilde{\boldsymbol{\theta}}}'-\dot{\tilde{\boldsymbol{\theta}}}\right)dA_{\rho'}dA_{\rho} \tag{34}$$

$$\mathcal{W}_{sup}(\mathcal{P}) = \int_{\mathcal{P}}\int_{\mathcal{A}\backslash\mathcal{P}}\left(\underline{\mathbf{S}}\cdot\left(\dot{\mathbf{u}}'_S + \frac{\dot{\tilde{\boldsymbol{\theta}}}'}{2}\times(\boldsymbol{\rho}-\boldsymbol{\rho}')\right) - \underline{\mathbf{S}}'\cdot\left(\dot{\mathbf{u}}_S + \frac{\dot{\tilde{\boldsymbol{\theta}}}}{2}\times(\boldsymbol{\rho}'-\boldsymbol{\rho})\right)\right)dA_{\rho'}dA_{\rho}$$

$$+ \int_{\mathcal{P}}\int_{\mathcal{A}\backslash\mathcal{P}}\left(\underline{\boldsymbol{\mu}}\cdot\dot{\tilde{\boldsymbol{\theta}}}' - \underline{\boldsymbol{\mu}}'\cdot\dot{\tilde{\boldsymbol{\theta}}}\right)dA_{\rho'}dA_{\rho} + \int_{\mathcal{P}}\mathbf{b}_S\cdot\dot{\mathbf{u}}_S\,dA_{\rho} + \int_{\mathcal{P}}\mathbf{l}_S\cdot\dot{\tilde{\boldsymbol{\theta}}}\,dA_{\rho} \tag{35}$$

Since no other source of energy is considered, the absorbed power relates to the internal energy $\mathcal{E}(\mathcal{P})$ as $\dot{\mathcal{E}}(\mathcal{P}) = \mathcal{W}_{abs}(\mathcal{P})$. Also the existence of an internal energy density $e$ follows from the form of integrals in Eqn. (34) which show the additive character of the internal energy (i.e., $\mathcal{E}(\mathcal{P}_1) + \mathcal{E}(\mathcal{P}_2) = \mathcal{E}(\mathcal{P}_1 + \mathcal{P}_2)$). The internal energy density $e$ therefore has the rate form,

$$\dot{e} = \int_{\mathcal{A}} \underline{\mathbf{S}} \cdot \left\{ (\dot{\mathbf{u}}'_S - \dot{\mathbf{u}}_S) - \frac{\dot{\tilde{\boldsymbol{\theta}}} + \dot{\tilde{\boldsymbol{\theta}}}'}{2} \times (\boldsymbol{\rho}' - \boldsymbol{\rho}) \right\} dA_{\boldsymbol{\rho}'} + \int_{\mathcal{A}} \underline{\boldsymbol{\mu}} \cdot \left( \dot{\tilde{\boldsymbol{\theta}}}' - \dot{\tilde{\boldsymbol{\theta}}} \right) dA_{\boldsymbol{\rho}'} \qquad (36)$$

Using the inner product of surface vectors defined as $\underline{\mathbf{A}} \bullet \underline{\mathbf{B}} = \int_{\mathcal{A}} \underline{\mathbf{A}} \langle \tilde{\boldsymbol{\xi}} \rangle \cdot \underline{\mathbf{B}} \langle \tilde{\boldsymbol{\xi}} \rangle dA_{\boldsymbol{\rho}'}$, the expression above reduces as,

$$\dot{e} == \underline{\mathbf{S}} \bullet \left( \underline{\dot{\mathbf{U}}} - \underline{\dot{\hat{\Theta}}} \times \underline{\boldsymbol{\rho}} \right) + \underline{\boldsymbol{\mu}} \bullet \underline{\dot{\Theta}} = \underline{\mathbf{S}} \bullet \underline{\dot{\mathbf{U}}}_{\hat{\Theta}} + \underline{\boldsymbol{\mu}} \bullet \underline{\dot{\Theta}}, \qquad (37)$$

$\underline{\mathbf{U}}_{\hat{\Theta}} = \underline{\mathbf{U}} - \underline{\hat{\Theta}} \times \underline{\boldsymbol{\rho}}$ is a composite state and its action on a curved bond $\tilde{\boldsymbol{\xi}}$ is given as

$$\underline{\mathbf{U}}_{\hat{\Theta}} \langle \tilde{\boldsymbol{\xi}} \rangle = (\mathbf{u}'_S - \mathbf{u}_S) - \frac{\tilde{\boldsymbol{\theta}} + \tilde{\boldsymbol{\theta}}'}{2} \times (\boldsymbol{\rho}' - \boldsymbol{\rho}).$$

Assuming the energy density functional $e$ to depend on $\underline{\mathbf{Y}}_{\hat{\Theta}}$ and $\underline{\Theta}$, the rate of $e$ would be given by the following expression.

$$\dot{e}(\underline{\mathbf{U}}_{\hat{\Theta}}, \underline{\Theta}) = e_{\underline{\mathbf{U}}_{\hat{\Theta}}} \bullet \underline{\dot{\mathbf{U}}}_{\hat{\Theta}} + e_{\underline{\Theta}} \bullet \underline{\dot{\Theta}}, \qquad (38)$$

where $e_{\underline{\mathbf{U}}_{\hat{\Theta}}}$ and $e_{\underline{\Theta}}$ are the Fréchet derivatives (Silling *et al.*, 2007) of $e$ with respect to $\underline{\mathbf{U}}_{\hat{\Theta}}$ and $\underline{\Theta}$ respectively. Comparison of (37) with (38) leads to the following constitutive relations (which are consistent with the second law of thermodynamics and also obtainable by the Coleman-Noll procedure; see Tadmor *et al.*, 2012).

$$\underline{\mathbf{S}} = e_{\underline{\mathbf{U}}_{\hat{\Theta}}} \quad \text{and} \quad \underline{\boldsymbol{\mu}} = e_{\underline{\Theta}}.$$

In order to write down a more explicit form of the constitutive relations, we opt for a constitutive correspondence framework similar to what is suggested in (Silling *et al.*, 2007) and make use of the classical constitution for linear elastic isotropic shell (see Eqn. (9) and (10)). In the following two subsections we introduce two non-local strain tensors derived from the PD kinematic states, show their correspondence to local strains and subsequently establish the PD constitution for the shell.

*3.3.1 Non-local strain measures and kinematic correspondence*

For continuously differentiable deformation and rotation fields on the base surface, the following identities hold.

$$\underline{\mathbf{U}}_{\hat{\Theta}}[\boldsymbol{\rho}]\langle\tilde{\boldsymbol{\xi}}\rangle = (\mathbf{u}'_S - \mathbf{u}_S) - \frac{\tilde{\boldsymbol{\theta}}' + \tilde{\boldsymbol{\theta}}}{2} \times (\boldsymbol{\rho}' - \boldsymbol{\rho})$$
$$= \left[(\tilde{\nabla}\mathbf{u}_S)^T - \mathbf{n}\otimes\mathbf{n}\times\tilde{\boldsymbol{\theta}}\right]\cdot\Delta\boldsymbol{\rho}^1 + O\left(\left\|\Delta\boldsymbol{\rho}^1\right\|^2\right) = \boldsymbol{\gamma}.\Delta\boldsymbol{\rho}^1 + O\left(\left\|\Delta\boldsymbol{\rho}^1\right\|^2\right) \quad (39)$$

$$\underline{\boldsymbol{\Theta}}[\boldsymbol{\rho}]\langle\tilde{\boldsymbol{\xi}}\rangle = \tilde{\boldsymbol{\theta}}' - \tilde{\boldsymbol{\theta}} = (\tilde{\nabla}\tilde{\boldsymbol{\theta}})^T\cdot\Delta\boldsymbol{\rho}^1 + O\left(\left\|\Delta\boldsymbol{\rho}^1\right\|^2\right) = \boldsymbol{\kappa}\cdot\Delta\boldsymbol{\rho}^1 + O\left(\left\|\Delta\boldsymbol{\rho}^1\right\|^2\right) \quad (40)$$

Here $\Delta\boldsymbol{\rho}^1$ is the projection of the bond $\tilde{\boldsymbol{\xi}}$ from $\boldsymbol{\rho}$ to $\boldsymbol{\rho}'$ to the tangent space of the base surface at $\boldsymbol{\rho}$ (i.e. the point whose position vector is $\boldsymbol{\rho}$) and is given by $\Delta\boldsymbol{\rho}^1 = \Delta q^1 \boldsymbol{\rho}_1 + \Delta q^2 \boldsymbol{\rho}_2$. $\Delta q^1$ and $\Delta q^2$ denote the changes in curvilinear coordinates $q^1$ and $q^2$ in moving from $\boldsymbol{\rho}$ to $\boldsymbol{\rho}'$.

We now define the following non-local approximations $\bar{\boldsymbol{\gamma}}$ and $\bar{\boldsymbol{\kappa}}$ to the local strain measures $\boldsymbol{\gamma}$ and $\boldsymbol{\kappa}$, with the former corresponding exactly to the local strain measures in case the deformation is homogeneous, i.e. yielding constant local strains.

$$\bar{\boldsymbol{\gamma}}(\underline{\mathbf{U}}_{\hat{\Theta}}) = \left[\int_{\tilde{\mathcal{H}}(\boldsymbol{\rho})}\tilde{\omega}(|\tilde{\boldsymbol{\xi}}|)\underline{\mathbf{U}}_{\hat{\Theta}}\langle\tilde{\boldsymbol{\xi}}\rangle\otimes\Delta\boldsymbol{\rho}^1 dA_{\boldsymbol{\rho}'}\right]\bar{\mathbf{K}}_S^{-1}, \quad \bar{\boldsymbol{\kappa}}(\underline{\boldsymbol{\Theta}}) = \left[\int_{\tilde{\mathcal{H}}(\boldsymbol{\rho})}\tilde{\omega}(|\tilde{\boldsymbol{\xi}}|)\underline{\boldsymbol{\Theta}}\langle\tilde{\boldsymbol{\xi}}\rangle\otimes\Delta\boldsymbol{\rho}^1 dA_{\boldsymbol{\rho}'}\right]\bar{\mathbf{K}}_S^{-1} \quad (41)$$

We define $\bar{\mathbf{K}}_S$, the shape tensor of the base surface, as $\bar{\mathbf{K}}_S = \int_{\tilde{\mathcal{H}}(\boldsymbol{\rho})}\tilde{\omega}(|\tilde{\boldsymbol{\xi}}|)\Delta\boldsymbol{\rho}^1\otimes\Delta\boldsymbol{\rho}^1 dA_{\boldsymbol{\rho}'}$. $\tilde{\omega}$ is a non-negative scalar influence function, defined over the base surface and satisfying $\tilde{\omega}(|\tilde{\boldsymbol{\xi}}|) > 0, \forall\tilde{\boldsymbol{\xi}}\in\tilde{\mathcal{H}}$.

*3.3.2 Constitutive correspondence*

We write the classical internal energy rate (see Eqn. (A.23) in Appendix I) in terms of the non-local strain measures defined in Eqn. (41) as,

$$\rho\dot{w}_S = \mathbf{S}\cdot\cdot\dot{\bar{\boldsymbol{\gamma}}} + \mathbf{M}\cdot\cdot\dot{\bar{\boldsymbol{\kappa}}} = \int_{\tilde{\mathcal{H}}(\boldsymbol{\rho})}\tilde{\omega}(|\tilde{\boldsymbol{\xi}}|)\mathbf{S}^T\bar{\mathbf{K}}_S^{-1}\Delta\boldsymbol{\rho}^1\cdot\underline{\dot{\mathbf{U}}}_{\hat{\Theta}}\langle\tilde{\boldsymbol{\xi}}\rangle dA_{\boldsymbol{\rho}'} + \int_{\tilde{\mathcal{H}}(\boldsymbol{\rho})}\tilde{\omega}(|\tilde{\boldsymbol{\xi}}|)\mathbf{M}^T\bar{\mathbf{K}}_S^{-1}\Delta\boldsymbol{\rho}^1\cdot\underline{\dot{\boldsymbol{\Theta}}}\langle\tilde{\boldsymbol{\xi}}\rangle dA_{\boldsymbol{\rho}'} \quad (42)$$

Symmetry of the shape tensor is utilised in writing the above equation. The double contraction operator $\cdot\cdot$ is defined as $(\mathbf{a}\otimes\mathbf{b})\cdot\cdot(\mathbf{c}\otimes\mathbf{d}) = (\mathbf{a}\cdot\mathbf{d})(\mathbf{b}\cdot\mathbf{c})$ where $\mathbf{a},\mathbf{b},\mathbf{c},\mathbf{d}$ are vectors in $\mathbb{R}^3$. Equating Eqn. (42) with Eqn. (37), the following PD constitutive relations are obtained.

$$\underline{\mathbf{S}}[\boldsymbol{\rho}]\langle\tilde{\boldsymbol{\xi}}\rangle = \tilde{\omega}(|\tilde{\boldsymbol{\xi}}|)\mathbf{S}^T\bar{\mathbf{K}}_S^{-1}\Delta\boldsymbol{\rho}^1, \quad \underline{\mathbf{M}}[\boldsymbol{\rho}]\langle\tilde{\boldsymbol{\xi}}\rangle = \tilde{\omega}(|\tilde{\boldsymbol{\xi}}|)\mathbf{M}^T\bar{\mathbf{K}}_S^{-1}\Delta\boldsymbol{\rho}^1 \quad (43)$$

$\mathbf{S}$ and $\mathbf{M}$ in the above expression are given by

$$\mathbf{S} = \lambda h \, \text{tr}\left(\mathbf{A} \cdot \frac{\overline{\boldsymbol{\gamma}} + \overline{\boldsymbol{\gamma}}^{\text{T}}}{2}\right)\mathbf{A} + 2\mu h\left(\mathbf{A} \cdot \frac{\overline{\boldsymbol{\gamma}} + \overline{\boldsymbol{\gamma}}^{\text{T}}}{2}\right) \text{ and}$$

$$\mathbf{M} = \frac{\lambda h^3}{12}\left((\mathbf{A}.\overline{\boldsymbol{\kappa}})^{\text{T}} - \mathbf{A}.\overline{\boldsymbol{\kappa}}\right) + \frac{\lambda h^3}{6}\left((\mathbf{A}.\overline{\boldsymbol{\kappa}})^{\text{T}} - \frac{1}{2}\text{tr}\left((\mathbf{A}.\overline{\boldsymbol{\kappa}})^{\text{T}}\right)\mathbf{A}\right).$$

With the assumption of plane stress, the Lamé parameters may be given in terms of Young's modulus $E$ and Poisson's ratio $\nu$ as $\lambda = \dfrac{E\nu}{1-\nu^2}$ and $\mu = \dfrac{E}{2(1+\nu)}$.

.

## 4. Numerical Illustration

We now provide the solutions to three illustrative problems on bending of shells under uniformly distributed transverse static loading. For the first example, we choose a special case, viz. a shell with zero curvature, i.e. a plate, followed by a second example on a singly curved cylindrical shell and final one on a doubly curved spherical shell.

For numerical implementation, the 2D base surface is discretized using a finite number of nodes in its reference configuration. Each node, a PD particle, is associated with a finite area. All the governing equations are then written for these PD particles along with Riemann-sum type approximations (summations over these PD particles) for the integral quantities in these equations (Eqns. (15), (16)). This discretization results in a set of algebraic equations with displacement and rotation at the different PD particles as unknowns, which are solved for after imposing boundary conditions. Details of numerical implementation of a typical PD algorithm may be found elsewhere, e.g. Breitenfeld *et al*. (2014).

The numerical values chosen here for different material parameters are arbitrary. We use them only for the purpose of demonstration. In order to limit the computational burden, we choose the dimensions of the shells to be small. For all the three problems, the PD solutions are compared with those through the classical shell model obtained using the finite element solver Abaqus®. For such a comparison to make sense, the convergence of solutions based on the PD shell theory is sought by refining the particle distribution whilst simultaneously decreasing the horizon size. In the limit of such an exercise, the PD solution should converge to the classical one.

*4.1. Elastic deformation of a thin plate*

As the first example, we consider a plate, a special case of a shell where the curvature tensor $\mathbf{B}$ of the base surface is zero. To reduce the computational overhead, plate dimensions are considered small (length 0.05 m, width 0.05 m and thickness 0.001 m). The plate is assumed to be clamped at two opposite edges (say, along the width) with the other two edges free and subjected to a uniform transverse load $10^6 \text{ N}/\text{m}^2$. The plate material is taken to have the following elastic properties: $E = 100 \text{ GPa}$ and $\nu = 0.3$.

As we intend to investigate the static response of the plate, the inertial forces i.e. the quantities on the left hand side of Eqn. (15) and (16) are considered to be zero, reducing them to equilibrium equations. The curvature $\mathbf{B}$ being zero, we get two sets of uncoupled equations, one involving the in-plane displacement $\tilde{\mathbf{u}}$ and the other transverse displacement $w\mathbf{n}$ and rotation $\tilde{\boldsymbol{\theta}}$. It enables to solve for two smaller systems of algebraic equations following PD discretization. Under the application of transverse load alone, we in fact get $\tilde{\mathbf{u}}$ to be zero and only solve the transverse deflection and rotations. In order to assess the correctness of solutions via the proposed PD model, we aim to match the result with the finite element solution of classical plate equation. To converge to the classical solution from the PD equations, we choose the radius of the horizon (which is a circular disc on the plane of plate base surface) to be $\delta = 1.1d$ where $d$ is the Euclidean distance between two closest PD particles. We take a uniform discretization of the plate base surface and refine it to get to the converged solution. Figure 3(a) shows the transverse deflection of the base surface, whereas Figure 3(b) presents solutions for different discretization levels along the mid width of the plate. This figure also reports the result from the classical PDE based plate equation.

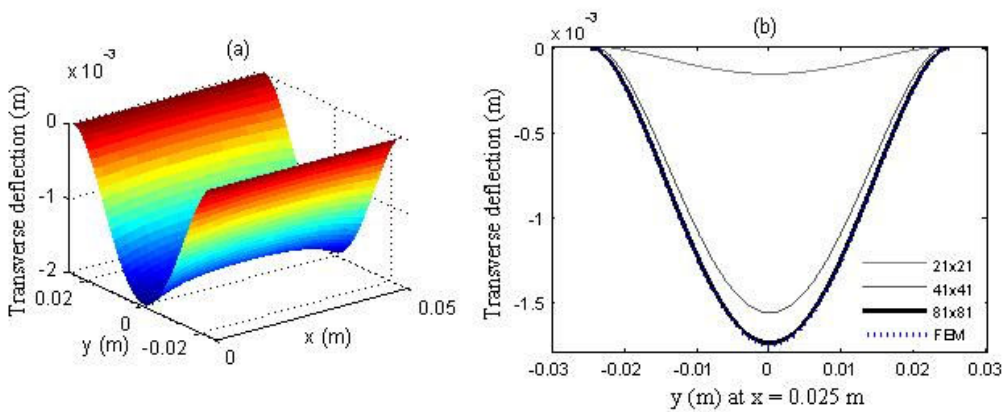

Figure 3. Transverse deflection of (a) mid surface and (b) along mid width

*4.2. Elastic deformation of a thin cylindrical shell*

For the second illustrative example, we consider a (singly curved) cylindrical shell. The curvature and thickness are chosen in such a way that $\|h\mathbf{B}\| \ll 1$, whereas the span and length may be chosen without any restriction. Once again, we consider small dimensions (radius of curvature 0.1m, span 0.05 m, length 0.05 m, thickness 0.001m) keeping the computational expense in mind. The shell is assumed to be clamped along the straight edges and free along the curved ones. Static deformation of the shell under a uniform transverse loading of $10^6$ N/m$^2$ is found by solving the PD model as well as the classical one. Note that, unlike plate, shell equations are fully coupled and have to be solved for the entire set of unknown displacements and rotations simultaneously. Figure 4(a) shows the deformed shape of the shell mid surface. We have used a deformation scale factor 10 to magnify the pattern of deflection. While the transverse deflection of the mid surface is reported in Figure 4(b), Figure 4(c) compares the PD and classical solutions along the mid length line. It also shows PD results for several discretization levels.

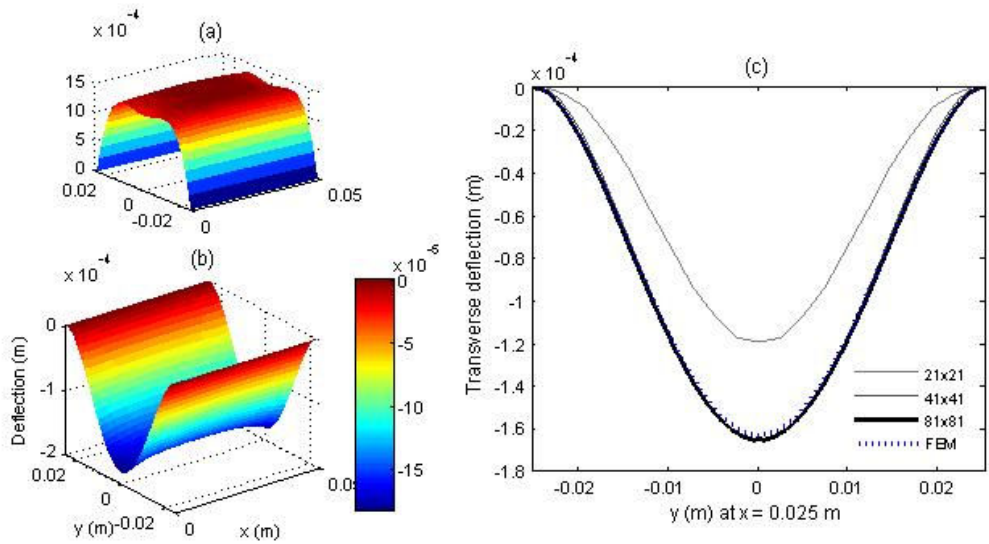

Figure 4. (a) Deformed shape of mid surface (displacement scale factor 10); Transverse deflection of (b) mid surface and (c) along mid length

*4.3. Elastic deformation of thin spherical shell*

We choose a portion of a spherical shell, a doubly curved surface, for the final illustrative example. Again following the restriction $\|h\mathbf{B}\| \ll 1$, the curvature and the thickness are chosen. Specifically, the dimensions considered are: radius of curvature 0.1m, thickness 0.001m, and the plan of the shell is $0.05 \times 0.05$ m$^2$. On all edges, clamped boundary condition is assumed and static deformation under uniform transverse loading of $10^6 \text{ N/m}^2$ is computed using both the PD equations as well as the classical theory. Deformed shape of the shell mid surface is presented in Figure 5(a), again with a deformation scale factor 10. Figure 5(b) shows the transverse deflection of the mid surface and Figure 5(c) compares the PD and classical solutions along the mid span line. PD results for several discretization levels are also shown here.

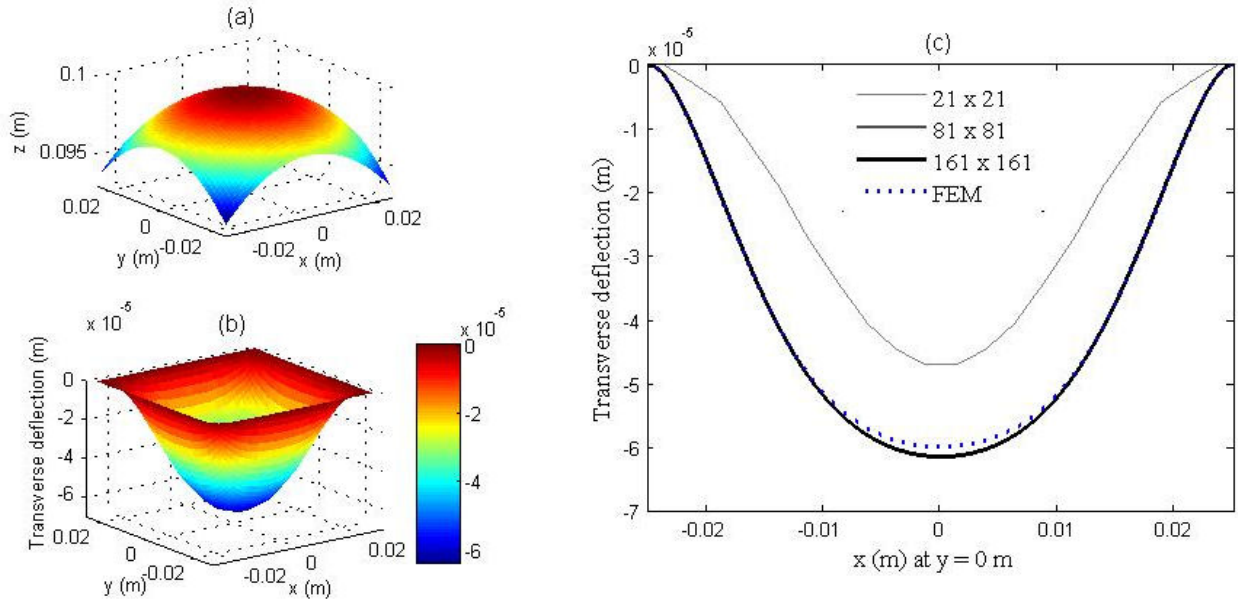

Figure 5. (a) Deformed shape of mid surface (displacement scale factor 10); Transverse deflection of (b) mid surface and (c) along mid span

## 4. Summary and concluding remarks

A peridynamic theory employing curved bonds and a surface based 2D formulation for linear elastic isotropic shells is presented. The proposed PD theory for thin shells has the advantage, among others, in circumventing the necessity of PD-based discretization along the thickness direction in a full-blown 3D model, an exercise fraught with numerical ill-conditioning and arduous computational overhead. The 2D formulation avoids the through-thickness discretization by analytically accounting for variations in the displacement and stress fields across the thickness dimension. The efficacy of the proposed model in analysing the elastic deformation of thin shells is demonstrated using a limited number of illustrative problems that include singly curved, doubly curved shells and a plate. Even though the presented scheme has limited applicability to only thin shells undergoing small elastic deformation, the PD shell framework, laid out in this work, lends an attractive modelling tool that may be useful in analysing brittle fracture of thin walled structures - a study left out of the current scope. Being restricted to only thin shells, the present shell formulation does not include any shear correction in the material constitution. An extension of the formulation to large elastic deformation cases and to those requiring shear correction, should also be of future interest.

**Appendix I. Classical formulation for small deformation of linear elastic shell**

The 2D classical equations of motion for a shell are derived from the 3D equations of motion through an appropriate integration over thickness. Shell constitutive relations for linear isotropic material are also arrived at from its 3D counterpart.

*A1. Geometry of shell and its base surface*

Shell is a 3D body that occupies a region $V_0 \in \mathbb{R}^3$ in its undeformed configuration and its boundary consists of three parts: upper and lower faces denoted by $S_+$ and $S_-$ respectively and the lateral surface $S_*$. The thickness $h$ of the shell is given by the normal distance between the surfaces $S_+$ and $S_-$. By the locus of the points equally distant ($h/2$) from both faces, the mid surface $S_0$ of the shell is defined. Considering $S_0$ as the base surface, the 2D equations would be derived in the subsequent sections.

The base surface $S_0$ is assumed to be sufficiently smooth. Any point $P$ on $S_0$ is represented by its position vector $\boldsymbol{\rho}$ with respect to a global Cartesian frame $\{o, \mathbf{e}_x, \mathbf{e}_y, \mathbf{e}_z\}$ as shown in Figure 2(b). In terms of the curvilinear surface coordinates $q^1$ and $q^2$, the position vector is given by $\boldsymbol{\rho} = \boldsymbol{\rho}(q^1, q^2)$. The tangent basis vectors $\boldsymbol{\rho}_1$ and $\boldsymbol{\rho}_2$ and its dual bases $\boldsymbol{\rho}^1$, $\boldsymbol{\rho}^2$ at $\boldsymbol{\rho}$ are defined as follows.

$$\boldsymbol{\rho}_\alpha = \frac{\partial \boldsymbol{\rho}}{\partial q^\alpha}, \qquad \boldsymbol{\rho}_\alpha \cdot \boldsymbol{\rho}^\beta = \delta_\alpha^\beta \qquad (\alpha, \beta = 1, 2) \tag{A.1}$$

In the above expression, $\delta_\alpha^\beta = 1$ for $\alpha = \beta$ and zero otherwise. A vector $\mathbf{n}$ normal to $S_0$ at $\boldsymbol{\rho}$ is defined as $\mathbf{n} = \dfrac{\boldsymbol{\rho}_1 \times \boldsymbol{\rho}_2}{|\boldsymbol{\rho}_1 \times \boldsymbol{\rho}_2|} = \dfrac{\boldsymbol{\rho}^1 \times \boldsymbol{\rho}^2}{|\boldsymbol{\rho}^1 \times \boldsymbol{\rho}^2|}$. At point $\boldsymbol{\rho}$, $(\boldsymbol{\rho}_1, \boldsymbol{\rho}_2, \mathbf{n})$ constitute a basis for $\mathbb{R}^3$ and the corresponding dual basis is given by $(\boldsymbol{\rho}^1, \boldsymbol{\rho}^2, \mathbf{n})$. The metric tensor $\mathbf{A}$ for the surface is defined as $\mathbf{A} = \boldsymbol{\rho}^\alpha \otimes \boldsymbol{\rho}_\alpha$. Another necessary information for the description of the base surface is its curvature tensor $\mathbf{B}$ which is defined as $\mathbf{B} = -\boldsymbol{\rho}^\alpha \otimes \dfrac{\partial \mathbf{n}}{\partial q^\alpha}$. Tensor $\mathbf{A}$ and $\mathbf{B}$ are respectively required for the definition first and second fundamental forms of $S_0$.

The characterization, as above, of the base surface allows representing any point $Q$ in the shell geometry given by a position vector $\mathbf{r}$ (with respect to the global Cartesian frame) in terms of the surface information as

$$\mathbf{r} = \mathbf{r}(q^1, q^2, z) = \boldsymbol{\rho}(q^1, q^2) + z\mathbf{n} \tag{A.2}$$

Here the normal through the point $Q$, denoted by $\mathbf{n}$, is assumed to have its base point at $P$, described through the position vector $\boldsymbol{\rho}$ on $S_0$. In other words, $Q$ is assigned coordinates $(q^1, q^2, z)$, with $\{q^1, q^2\} \in S_0$ and $z$ the third coordinate along $\mathbf{n}$. It may be noted that for points in the shell geometry, we have $-h/2 \leq z \leq h/2$ and $z$ is positive along the direction of $\mathbf{n}$. A set of tangent bases at point $Q$ and their relations with the tangent bases at point $P$ on the surface (see Eqn. (A.1)) is given below.

$$\mathbf{r}_\alpha = \frac{\partial \mathbf{r}}{\partial q^\alpha} = (\mathbf{A} - z\mathbf{B}) \cdot \boldsymbol{\rho}_\alpha, \qquad \mathbf{r}_3 = \frac{\partial \mathbf{r}}{\partial z} = \mathbf{n} \tag{A.3}$$

The tensor $(\mathbf{A} - z\mathbf{B})$ in the first of the expressions above is known as shift tensor and is useful in describing tensorial quantities defined at any point of the shell in terms of the surface basis vectors. The associated dual bases at point $Q$ have the following form.

$$\mathbf{r}^\alpha = (\mathbf{A} - z\mathbf{B})^{-1} \cdot \boldsymbol{\rho}^\alpha, \qquad \mathbf{r}^3 = \mathbf{n} \tag{A.4}$$

The shift tensor $(\mathbf{A} - z\mathbf{B})$ is singular as a 3D tensor; therefore, its inverse is obtainable by considering $(\mathbf{A} - z\mathbf{B})$ as an operator on the 2D tangent space of the base surface leading to the identity (A.5) below. See (Lebedev *et al.*, 2010) for details.

$$(\mathbf{A} - z\mathbf{B}) \cdot (\mathbf{A} - z\mathbf{B})^{-1} = (\mathbf{A} - z\mathbf{B})^{-1} \cdot (\mathbf{A} - z\mathbf{B}) = \mathbf{A} \tag{A.5}$$

For spatial differentiation, the definition of the gradient operator is taken as

$$\nabla = \mathbf{r}^\alpha \otimes \frac{\partial}{\partial q^\alpha} + \mathbf{n} \otimes \frac{\partial}{\partial z} = (\mathbf{A} - z\mathbf{B})^{-1} \cdot \boldsymbol{\rho}^\alpha \otimes \frac{\partial}{\partial q^\alpha} + \mathbf{n} \otimes \frac{\partial}{\partial z} = (\mathbf{A} - z\mathbf{B})^{-1} \cdot \tilde{\nabla} + \mathbf{n} \otimes \frac{\partial}{\partial z} \tag{A.6}$$

where $\tilde{\nabla} = \boldsymbol{\rho}^\alpha \otimes \frac{\partial}{\partial q^\alpha}$ is the gradient operator on the base surface $S_0$. Eqn. (A.6) holds when the gradient of a vector or a higher order tensor field has to be defined; however in case of a scalar field, the necessary definition for the gradient operator would be similar to Eqn. (A.6) with no dyadic product.

*A2. Kinematic hypothesis and displacement field*

Recourse to the hypothesis method of shell modelling is taken in this article. In the context of the shell kinematics, any line segment along the normal to the undeformed base surface is hypothesised to remain linear and unstretched upon deformation. Normality of this deformed line segment to the deformed base surface is not assumed. These hypotheses allow writing the 3D displacement of any point (say $Q$) in the shell in the following form.

$$\mathbf{u}(q^1, q^2, z) = \tilde{\mathbf{u}}(q^1, q^2) + w(q^1, q^2)\mathbf{n} + \tilde{\boldsymbol{\theta}}(q^1, q^2) \times z\mathbf{n} \tag{A.7}$$

In the above expression, $\tilde{\mathbf{u}} + w\mathbf{n}$ represents the displacement vector field of the base surface $S_0$, wherein $\tilde{\mathbf{u}} = \tilde{u}_1 \boldsymbol{\rho}^1 + \tilde{u}_2 \boldsymbol{\rho}^2$ is the projection of this vector field onto the tangent planes at the respective base points and $w\mathbf{n}$ the projection normal to the tangent planes. $\tilde{\boldsymbol{\theta}} = \tilde{\theta}_1 \boldsymbol{\rho}^1 + \tilde{\theta}_2 \boldsymbol{\rho}^2$ is a small rotation vector field defined at the base surface. As may be seen, the rotation does not include components in the normal direction; therefore the drilling rotation is not considered in this formulation. It may be noted that the hypotheses here closely relate to Reissner's theory of shell (Reissner, 1941).

Following the definition given in Eqn. (A.6), we may write the spatial gradient of the displacement field $\mathbf{u}$ as

$$\nabla \mathbf{u} = \nabla\left(\tilde{\mathbf{u}} + w\mathbf{n} + \tilde{\boldsymbol{\theta}} \times z\mathbf{n}\right) = \left(\mathbf{A} - z\mathbf{B}\right)^{-1} \cdot \left(\tilde{\nabla}\tilde{\mathbf{u}} \cdot \mathbf{A} + \left(\mathbf{B} \cdot \mathbf{u}\right) \otimes \mathbf{n} - w\mathbf{B}\right) + \nabla\left(\tilde{\boldsymbol{\theta}} \times z\mathbf{n}\right) \tag{A.8}$$

which will be utilized subsequently in identifying surface strain measures and also in writing energy balance equations.

*A3. Equations of motion*

In order to derive the shell equations of motion in terms of the surface quantities, we start with writing the mechanical version of the first law of thermodynamics for an 'arbitrary' volume $V$ in the reference configuration of the 3D shell. The arbitrariness in $V$ is entirely due to that in the area of the base surface it encompasses; however $V$ extends over the entire thickness (see Figure 2(a) for the volume element considered). Let the surface that encloses $V$ be denoted by $\partial V$. Invoking invariance of this law under superimposed rigid translation and rotation followed by the through-thickness integration would lead to the set of equations of motion.

Mechanical version of the first law of thermodynamics (i.e., when heat effects are not considered) states that the rate of increase of the total energy (kinetic and internal) of any portion of the body equals the rate of work done on that portion by the body force and surface traction. Mathematically this translates to the following.

$$\frac{d}{dt}\left(\mathcal{U} + \mathcal{K}\right) = \mathcal{L} \tag{A.9}$$

Here $\mathcal{U}$, $\mathcal{K}$ and $\mathcal{L}$ denote respectively the internal energy, kinetic energy and rate of work done by the external forces and are defined for the arbitrary volume $V$ as

$$\mathcal{U} = \int_V \rho w \, dV, \quad \mathcal{K} = \int_V \rho \mathbf{v} \cdot \mathbf{v} \, dV, \quad \mathcal{L} = \int_V \rho \mathbf{f} \cdot \mathbf{v} \, dV + \int_{\partial V} \mathbf{t} \cdot \mathbf{v} \, dV \tag{A.10}$$

$\rho, w, \mathbf{v}, \mathbf{f}$ and $\mathbf{t}$ are respectively the mass density, specific internal energy, velocity vector field, body force vector and Cauchy traction. The mass density $\rho$ is assumed not to vary along the thickness. Velocity vector for the shell is obtainable from Eqn. (A.7) upon taking the material derivative. The traction vector satisfies $\mathbf{t} = \mathbf{n}_{\partial V} \cdot \boldsymbol{\sigma}$ where $\boldsymbol{\sigma}$ is the Cauchy stress tensor and $\mathbf{n}_{\partial V}$ the normal to the surface $\partial V$. Substituting for $\mathbf{t}$ and applying divergence theorem, the rate of work takes the form $\mathcal{L} = \int_V \rho \mathbf{f} \cdot \mathbf{v} \, dV + \int_V (\mathbf{v} \cdot \nabla \cdot \boldsymbol{\sigma} + \boldsymbol{\sigma} \cdot \cdot \nabla \mathbf{v}) \, dV$. Eqn. (A.9) may then be recast as follows.

$$\int_V \rho \dot{w} dV = \int_V \mathbf{v} \cdot (\nabla \cdot \boldsymbol{\sigma} + \rho \mathbf{f} - \rho \dot{\mathbf{v}}) dV + \int_V \boldsymbol{\sigma} \cdot \cdot \nabla \mathbf{v} dV$$
$$= \int_V \mathbf{v} \cdot \left( \frac{1}{\sqrt{g}} \left( \frac{\partial(\sqrt{g}\boldsymbol{\sigma}^\alpha)}{\partial q^\alpha} + \frac{\partial(\sqrt{g}\boldsymbol{\sigma}^3)}{\partial z} \right) + \rho \mathbf{f} - \rho \dot{\mathbf{v}} \right) dV + \int_V \boldsymbol{\sigma} \cdot \cdot \nabla \mathbf{v} dV \quad (A.11)$$

The second line in (A.11) may be obtained by expanding $\nabla \cdot \boldsymbol{\sigma}$ in terms of the basis vectors appearing in Eqn. (A.3) and (A.4). $g = aG^2$ is the determinant of a matrix whose $(i,j)$ th element is $\mathbf{r}_i \cdot \mathbf{r}_j$ $(i,j = 1,2,3)$. $a = a_{11}a_{22} - a_{12}^2$ where $a_{\alpha\beta}$ are the covariant components of the metric tensor $\mathbf{A}$ and $G = \det(\mathbf{A} - z\mathbf{B})$.

*Invariance under rigid translation:* The statement of the first law of thermodynamics must be invariant with respect to rigid motions of the body. We first consider rectilinear uniform motion, and substitute $\mathbf{v} + \mathbf{b}$ for $\mathbf{v}$ in Eqn. (A.11) to obtain the following.

$$\int_V \rho \dot{w} dV = \int_V (\mathbf{v} + \mathbf{b}) \cdot \left( \frac{1}{\sqrt{g}} \left( \frac{\partial(\sqrt{g}\boldsymbol{\sigma}^\alpha)}{\partial q^\alpha} + \frac{\partial(\sqrt{g}\boldsymbol{\sigma}^3)}{\partial z} \right) + \rho \mathbf{f} - \rho \dot{\mathbf{v}} \right) dV + \int_V \boldsymbol{\sigma} \cdot \cdot \nabla \mathbf{v} dV \quad (A.12)$$

$\mathbf{b}$ is a constant vector in $\mathbb{R}^3$, $\dot{\mathbf{b}} = 0$ and $\nabla \mathbf{b} = 0$ and thus Eqn. (A.12) follows from Eqn. (A.11). The volume element $dV = \sqrt{a}G dq^1 dq^2 dz$. Subtracting Eqn. (A.11) from Eqn. (A.12) we arrive at

$$\int_V \mathbf{b} \cdot \left( \frac{1}{\sqrt{g}} \left( \frac{\partial(\sqrt{g}\boldsymbol{\sigma}^\alpha)}{\partial q^\alpha} + \frac{\partial(\sqrt{g}\boldsymbol{\sigma}^3)}{\partial z} \right) + \rho \mathbf{f} - \rho \dot{\mathbf{v}} \right) dV = 0 \quad (A.13)$$

Writing the volume integral as $\int_V dV = \int_A \int_{-h/2}^{h/2} G\,dz\,dA$ where $A$ represents the area of the base surface enclosed within $V$ and $dA = \sqrt{a}\,dq^1 dq^2$, Eqn. (A.13) may be rewritten as

$$\mathbf{b} \cdot \int_A \left( \frac{1}{\sqrt{a}} \frac{\partial \left( \sqrt{a} [\![\boldsymbol{\sigma}^\alpha]\!] \right)}{\partial q^\alpha} + G_+ \mathbf{t}_+^0 - G_- \mathbf{t}_-^0 + [\![\rho \mathbf{f}]\!] - [\![\rho \dot{\mathbf{v}}]\!] \right) dA = 0 \qquad (A.14)$$

in which $\mathbf{t}_+^0$ and $\mathbf{t}_-^0$ are the prescribed surface traction on $S_+$ and $S_-$ respectively and $[\![\bullet]\!] = \int_{-h/2}^{h/2} (\bullet) G\,dz$.

Eqn. (A.14) must be satisfied for every constant vector $\mathbf{b}$ and for any arbitrary area $A$; therefore upon localization we obtain the following.

$$\frac{1}{\sqrt{a}} \frac{\partial \left( \sqrt{a} [\![\boldsymbol{\sigma}^\alpha]\!] \right)}{\partial q^\alpha} + G_+ \mathbf{t}_+^0 - G_- \mathbf{t}_-^0 + [\![\rho \mathbf{f}]\!] - [\![\rho \dot{\mathbf{v}}]\!] = \mathbf{0} \qquad (A.15)$$

We now introduce the stress tensor $\mathbf{S}$ defined on the base surface as $\mathbf{S} = \boldsymbol{\rho}_\alpha \otimes [\![\boldsymbol{\sigma}^\alpha]\!] = [\![(\mathbf{A} - z\mathbf{B})^{-1} \cdot \boldsymbol{\sigma}]\!]$ and recast Eqn. (A.15) as

$$\tilde{\nabla} \cdot \mathbf{S} + \mathbf{q} = [\![\rho \dot{\mathbf{v}}]\!] \qquad (A.16)$$

where $\mathbf{q} = G_+ \mathbf{t}_+^0 - G_- \mathbf{t}_-^0 + [\![\rho \mathbf{f}]\!]$ acts as a 'body' force vector.

*Invariance under rigid rotation:* Eqn. (A.11) must also be invariant with respect to rotation of the body with constant angular velocity $\boldsymbol{\omega}$. Assuming the angular motion to be with respect to an axis passing through the origin, we replace $\mathbf{v}$ by $\mathbf{v} + \boldsymbol{\omega} \times \mathbf{r}$ in Eqn. (A.11) and then subtract Eqn. (A.11) to obtain the following.

$$\int_V (\boldsymbol{\omega} \times \mathbf{r}) \cdot \left( \frac{1}{\sqrt{g}} \left( \frac{\partial \left( \sqrt{g} \boldsymbol{\sigma}^\alpha \right)}{\partial q^\alpha} + \frac{\partial \left( \sqrt{g} \boldsymbol{\sigma}^3 \right)}{\partial z} \right) + \rho \mathbf{f} - \rho \dot{\mathbf{v}} \right) dV + \int_V \boldsymbol{\sigma} \cdot \cdot (\boldsymbol{\omega} \times \nabla \mathbf{r}) dV = 0 \qquad (A.17)$$

Using symmetry of $\boldsymbol{\sigma}$ and the identity $(\boldsymbol{\omega} \times \nabla \mathbf{r})^\mathrm{T} = -(\boldsymbol{\omega} \times \nabla \mathbf{r})$, we get $\boldsymbol{\sigma} \cdot \cdot (\boldsymbol{\omega} \times \nabla \mathbf{r}) = 0$. Eqn. (A.17) may then be simplified using Eqn. (A.2) and Eqn. (A.15) as

$$\boldsymbol{\omega} \cdot \int_A \int_{-h/2}^{h/2} \left( \frac{1}{\sqrt{a}} \frac{\partial \left( z\mathbf{n} \times \sqrt{a} G \boldsymbol{\sigma}^\alpha \right)}{\partial q^\alpha} - \frac{\partial \mathbf{n}}{\partial q^\alpha} \times G z \boldsymbol{\sigma}^\alpha + z \frac{\partial \left( \mathbf{n} \times G \boldsymbol{\sigma}^3 \right)}{\partial z} + \rho G z \mathbf{n} \times \mathbf{f} - \rho G z \mathbf{n} \times \dot{\mathbf{v}} \right) dz\,dA = 0 \, (A.18)$$

The above must hold for any constant $\boldsymbol{\omega}$ and arbitrary $A$. This allows localization of Eqn. (A.18), leading to

$$\frac{1}{\sqrt{a}}\frac{\partial\left(\sqrt{a}[\![z\mathbf{n}\times\boldsymbol{\sigma}^\alpha]\!]\right)}{\partial q^\alpha}+\boldsymbol{\rho}_\alpha\times[\![\boldsymbol{\sigma}^\alpha]\!]+\frac{h}{2}G_+\mathbf{n}\times\mathbf{t}_+^0-\frac{h}{2}G_-\mathbf{n}\times\mathbf{t}_-^0+[\![\rho z\mathbf{n}\times\mathbf{f}]\!]-[\![\rho z\mathbf{n}\times\dot{\mathbf{v}}]\!]=\mathbf{0} \qquad (A.19)$$

Here the identities $\dfrac{\partial\mathbf{n}}{\partial q^\alpha}\times Gz\boldsymbol{\sigma}^\alpha=-G\left(\mathbf{n}\times\boldsymbol{\sigma}^3+\boldsymbol{\rho}_\alpha\times\boldsymbol{\sigma}^\alpha\right)$ obtained using Eqn. (A.3) and

$$z\frac{\partial\left(G\mathbf{n}\times\boldsymbol{\sigma}^3\right)}{\partial z}=\frac{\partial\left(Gz\mathbf{n}\times\boldsymbol{\sigma}^3\right)}{\partial z}-G\mathbf{n}\times\boldsymbol{\sigma}^3$$

are made used of. Introducing the moment tensor $\mathbf{M}$ defined on the base surface as $\mathbf{M}=\boldsymbol{\rho}_\alpha\otimes[\![z\mathbf{n}\times\boldsymbol{\sigma}^\alpha]\!]=-[\![(\mathbf{A}-z\mathbf{B})^{-1}\cdot z\boldsymbol{\sigma}\times\mathbf{n}]\!]$ and defining the vectorial invariant of $\mathbf{S}$ as $\mathbf{S}_\times=\boldsymbol{\rho}_\alpha\times[\![\boldsymbol{\sigma}^\alpha]\!]$, Eqn. (A.19) takes the form

$$\tilde{\nabla}\cdot\mathbf{M}+\mathbf{S}_\times+\mathbf{m}=[\![\rho z\mathbf{n}\times\dot{\mathbf{v}}]\!] \qquad (A.20)$$

Here $\mathbf{m}=\dfrac{h}{2}G_+\mathbf{n}\times\mathbf{t}_+^0-\dfrac{h}{2}G_-\mathbf{n}\times\mathbf{t}_-^0+[\![\rho z\mathbf{n}\times\mathbf{f}]\!]$ acts as a 'body' couple.

The set of equations given, viz. Eqn. (A.16) and (A.20), represents the surface based equations of motion for the shell.

*A4. Conjugate surface strain measures*

In section A2, an expression for 3D displacement field of the shell in terms of the surface displacement and rotation is given (see Eqn. (A7)). However, development of the shell constitutive model would require defining the surface strain measures in terms of these displacement and rotation fields. In this section we identify the surface strain measures as quantities conjugate to the surface stress tensor $\mathbf{S}$ and moment tensor $\mathbf{M}$. Towards this, we rewrite Eqn. (A.11) as

$$\begin{aligned}\int_V \rho\dot{u}\dot{v}dV &= \int_A\left(\dot{\tilde{\mathbf{u}}}+\dot{w}\mathbf{n}\right)\cdot\int_{-h/2}^{h/2}\left(\frac{1}{\sqrt{a}}\left(\frac{\partial\left(\sqrt{a}G\boldsymbol{\sigma}^\alpha\right)}{\partial q^\alpha}+\frac{\partial\left(\sqrt{g}\boldsymbol{\sigma}^3\right)}{\partial z}\right)+\rho G\mathbf{f}-\rho G\dot{\mathbf{v}}\right)dzdA \\ &\quad +\int_A\dot{\tilde{\boldsymbol{\theta}}}\cdot\int_{-h/2}^{h/2}z\mathbf{n}\times\left(\frac{1}{\sqrt{a}}\left(\frac{\partial\left(\sqrt{a}G\boldsymbol{\sigma}^\alpha\right)}{\partial q^\alpha}+\frac{\partial\left(\sqrt{g}\boldsymbol{\sigma}^3\right)}{\partial z}\right)+\rho G\mathbf{f}-\rho G\dot{\mathbf{v}}\right)dzdA+\int_V\boldsymbol{\sigma}\cdot\cdot\nabla\mathbf{v}dV\end{aligned}$$

(A.21)

The first two terms on the right hand side of the above expression are identically zero owing to equations of motion given in Eqn. (A.15) and (A.19). This simplifies Eqn. (A.21) as $\int_V \rho \dot{w} dV = \int_V \boldsymbol{\sigma} \cdot\cdot \nabla \mathbf{v} dV$ or alternatively

$$\int_A \rho \dot{w}_S dA = \int_A \int_{-h/2}^{h/2} \boldsymbol{\sigma} \cdot\cdot \nabla \mathbf{v} G dz dA \quad (A.22)$$

where $w_S$, an energy density for the surface and defined as $w_S = [\![w]\!]$, is introduced. The assumption of $\rho$ not varying along $z$ is used to obtain the left hand side of Eqn. (A.22). Upon localization, Eqn. (A.22) leads to

$$\rho \dot{w}_S = \int_{-h/2}^{h/2} \boldsymbol{\sigma} \cdot\cdot \nabla \mathbf{v} G dz = \mathbf{S} \cdot\cdot \left( \left( \tilde{\nabla} \dot{\tilde{\mathbf{u}}} \right)^T + \left( \tilde{\nabla}(\dot{w}\mathbf{n}) \right)^T - \mathbf{n} \otimes \mathbf{n} \times \dot{\tilde{\boldsymbol{\theta}}} \right) + \mathbf{M} \cdot\cdot \left( \tilde{\nabla} \dot{\tilde{\boldsymbol{\theta}}} \right)^T = \mathbf{S} \cdot\cdot \dot{\boldsymbol{\gamma}} + \mathbf{M} \cdot\cdot \dot{\boldsymbol{\kappa}} \quad (A.23)$$

Here $\nabla \mathbf{v} = (\mathbf{A} - z\mathbf{B})^{-1} \cdot \left( \tilde{\nabla} \dot{\tilde{\mathbf{u}}} \cdot \mathbf{A} + (\mathbf{B} \cdot \dot{\tilde{\mathbf{u}}}) \otimes \mathbf{n} - \dot{w}\mathbf{B} \right) + \nabla \left( \dot{\tilde{\boldsymbol{\theta}}} \times z\mathbf{n} \right)$ is made use of. From Eqn. (A.23), the surface strain measures $\boldsymbol{\gamma}$ and $\boldsymbol{\kappa}$ are identified as

$$\boldsymbol{\gamma} = \left( \tilde{\nabla} \tilde{\mathbf{u}} \right)^T + \left( \tilde{\nabla}(w\mathbf{n}) \right)^T - \mathbf{n} \otimes \mathbf{n} \times \tilde{\boldsymbol{\theta}}, \qquad \boldsymbol{\kappa} = \left( \tilde{\nabla} \tilde{\boldsymbol{\theta}} \right)^T \quad (A.24)$$

*A5. Constitutive relations*

Starting with the 3D constitutive relations in terms of 3D stress and strain measures, the 2D constitutive equations are derived here. For the 3D model, a linear elastic material as given below is considered.

$$\boldsymbol{\sigma} = \lambda \mathrm{tr}(\mathbf{E})\mathbf{I} + 2\mu \mathbf{E} \quad (A.25)$$

Here $\lambda$ and $\mu$ are the Lamé constants and $\mathbf{E} = \left( \nabla \mathbf{u} + (\nabla \mathbf{u})^T \right) / 2$ is the small strain measure.

For the shell formulation here, we focus only on those satisfying $\|h\mathbf{B}\| \ll 1$. Upon imposing this condition, following approximation to $\mathbf{S}$ is arrived at.

$$\mathbf{S} = \int_{-h/2}^{h/2} (\mathbf{A} - z\mathbf{B})^{-1} \cdot \boldsymbol{\sigma} G dz \approx \int_{-h/2}^{h/2} \mathbf{A} \cdot \boldsymbol{\sigma} dz \quad (A.26)$$

Using Eqn. (A.8) and (A.25), we may recast Eqn. (A.26) as

$$\mathbf{S} = \lambda h \mathrm{tr} \left( \frac{\mathbf{A} \cdot \left( \tilde{\nabla} \tilde{\mathbf{u}} \right)^T + \tilde{\nabla} \tilde{\mathbf{u}} \cdot \mathbf{A}}{2} - w\mathbf{B} \right) \mathbf{A} + 2\mu h \left( \frac{\mathbf{A} \cdot \left( \tilde{\nabla} \tilde{\mathbf{u}} \right)^T + \tilde{\nabla} \tilde{\mathbf{u}} \cdot \mathbf{A}}{2} + \frac{(\mathbf{B} \cdot \tilde{\mathbf{u}}) \otimes \mathbf{n}}{2} + \frac{\tilde{\nabla} w \otimes \mathbf{n}}{2} - w\mathbf{B} + \frac{\tilde{\boldsymbol{\theta}} \times \mathbf{n} \otimes \mathbf{n}}{2} \right)$$

$$= \lambda h \, \mathrm{tr} \left( \mathbf{A} \cdot \frac{\boldsymbol{\gamma} + \boldsymbol{\gamma}^T}{2} \right) \mathbf{A} + 2\mu h \left( \mathbf{A} \cdot \frac{\boldsymbol{\gamma} + \boldsymbol{\gamma}^T}{2} \right) \quad (A.27)$$

Similarly, from the definition of $\mathbf{M}$, the following may be obtained.

$$\begin{aligned}\mathbf{M} &= -\int_{-h/2}^{h/2} \left(\mathbf{A} - z\mathbf{B}\right)^{-1} \cdot z\boldsymbol{\sigma} \times \mathbf{n} G dz \approx -\int_{-h/2}^{h/2} \mathbf{A} \cdot z\boldsymbol{\sigma} \times \mathbf{n} dz \\ &= \frac{\lambda h^3}{12}\left(\left(\mathbf{A}.\boldsymbol{\kappa}\right)^{\mathrm{T}} - \mathbf{A}.\boldsymbol{\kappa}\right) + \frac{\lambda h^3}{6}\left(\left(\mathbf{A}.\boldsymbol{\kappa}\right)^{\mathrm{T}} - \frac{1}{2}\mathrm{tr}\left(\left(\mathbf{A}.\boldsymbol{\kappa}\right)^{\mathrm{T}}\right)\mathbf{A}\right)\end{aligned} \tag{A.28}$$